\documentclass[12pt,twocolumn]{emulateapj}

\tightenlines

\usepackage{ulem} 
\usepackage{xcolor}
\usepackage{graphicx}
\usepackage{amsmath}
\usepackage{enumerate}
\usepackage{float}
\usepackage[caption = false]{subfig}

\newcount\listnorom
\newcommand\listromanDE{\global\advance \listnorom by 1
{\lowercase\expandafter{(\romannumeral\listnorom)}\ }}

\newcount\listnumber
\newcommand\listDE{\global\advance \listnumber by 1
{\lowercase\expandafter{(\number\listnumber)}\ }}

\def\lsim{\raise0.3ex
  \hbox{$<$\kern-0.75em\raise-1.1ex\hbox{$\sim$}}\,}
\def\gsim{\raise0.3ex
  \hbox{$>$\kern-0.75em\raise-1.1ex\hbox{$\sim$}}\,}

\newcommand{\be}{\begin{eqnarray}}
\newcommand{\ee}{\end{eqnarray}}

\newcommand{\mrm}{\mathrm}

\newcount\Inum
\newcount\IInum
\newcount\IIInum
\newcount\IVnum
\def\I{\global\multiply\IInum by 0 \global\multiply\IIInum by 0
            \global\multiply\IVnum by 0 \global\advance \Inum by 1
            {\the\Inum. }}
\def\II{\global\multiply\IIInum by 0\global\multiply\IVnum by 0
       \global\advance \IInum by 1 {\the\Inum.\the\IInum. }}
\def\III{\global\multiply\IVnum by 0\global\advance \IIInum by 1
            {\the\Inum.\the\IInum.\the\IIInum. }}
\def\IV{\global\advance \IVnum by 1
            {\the\IVnum. }}

\begin{document}

\title{Modeling Bright $\gamma$-ray and Radio Emission at Fast Cloud Shocks} 

\author{
Shiu-Hang Lee\altaffilmark{1,4,5},
Daniel J. Patnaude \altaffilmark{2},
John C. Raymond \altaffilmark{2},
Shigehiro Nagataki \altaffilmark{4},
Patrick O. Slane \altaffilmark{2} and
Donald C. Ellison \altaffilmark{3} 
}

\altaffiltext{1}{Institute of Space and Astronautical Science, Japan Aerospace Exploration Agency, 
3-1-1 Yoshinodai, Chuo-ku, Sagamihara, Kanagawa, 252-5210, Japan; 
slee@astro.isas.jaxa.jp}
\altaffiltext{2}{Harvard-Smithsonian Center for Astrophysics, 
60 Garden Street, Cambridge, MA 02138, U.S.A.;
slane@cfa.harvard.edu; dpatnaude@cfa.harvard.edu; jraymond@cfa.harvard.edu}
\altaffiltext{3}{Physics Department, North Carolina State
University, Box 8202, Raleigh, NC 27695, U.S.A.;
don\_ellison@ncsu.edu}
\altaffiltext{4}{RIKEN, Astrophysical Big Bang Laboratory and Interdisciplinary Theoretical Science Research Group, 
2-1 Hirosawa, Wako, Saitama 351-0198, Japan; 
shigehiro.nagataki@riken.jp}
\altaffiltext{5}{JAXA International Top Young Fellow}

\begin{abstract}
Recent observations by the Large Area Telescope (LAT) onboard the \textit{Fermi} satellite have revealed bright $\gamma$-ray emission from middle-aged supernova remnants (SNRs) inside our Galaxy. These remnants, which also possess bright non-thermal radio shells, are often found to be interacting directly with surrounding gas clouds. We explore the non-thermal emission mechanism at these dynamically evolved SNRs by constructing a hydrodynamical model. Two scenarios of particle acceleration, either a re-acceleration of Galactic cosmic rays (CRs) or an efficient nonlinear diffusive shock acceleration (NLDSA) of particles injected from downstream, are considered. Using parameters inferred from observations, our models are contrasted with the observed spectra of SNR W44. For the re-acceleration case, we predict a significant enhancement of radio and GeV emission as the SNR undergoes a transition into the radiative phase. If sufficiently strong magnetic turbulence is present in the molecular cloud, the re-acceleration scenario can explain the observed broadband spectral properties. The NLDSA scenario also succeeds in explaining the $\gamma$-ray spectrum but fails to reproduce the radio spectral index. Efficient NLDSA also results in a significant post-shock non-thermal pressure that limits the compression during cooling and prevents the formation of a prominent dense shell. Some other interesting differences between the two models in hydrodynamical behavior and resulting spectral features are illustrated. 
\end{abstract}


\section{Introduction}
Middle-aged SNRs have been receiving much attention recently thanks to the discovery of luminous $\gamma$-ray emission at their shells and sometimes close vicinities \citep[e.g.,][]{AbdoEtalW51C2009, AbdoEtalW282010, AbdoEtalW442010, AbdoEtalIC4432010,cs10}. The latest analysis of accumulated \textit{Fermi} LAT data on the SNRs W44 and IC 443 by \citet{Ackermann2013} has clearly uncovered a kinematic cutoff at around 250~MeV in their $\gamma$-ray spectra, a signature spectral characteristic that strongly supports the hadronic origin of the $\gamma$-rays (i.e. production of neutral pions through proton-proton interactions and their subsequent decay into photon pairs), and thus providing the long-awaited evidence for the acceleration of protons at SNR shocks. Earlier measurements by \textit{AGILE} have also shown evidences for such a signature \citep{Tavani2010, Giuliani2011}.
At the same time, however, the observed bright $\gamma$-ray emission raises new challenges to the conventional DSA theory of particle acceleration at astrophysical collisionless shocks. First of all, these  dynamically evolved remnants have slow shocks and hence are not normally expected to act as efficient particle accelerators like their younger cousins. This seems incongruent with the very high $\gamma$-ray luminosities inferred from observations. Moreover, these $\gamma$-rays that are concentrated in the GeV energy band possess a characteristic spectral shape that deviates from a simple power-law with an exponential cutoff; rather, it points to the existence of a peculiar spectral break in the underlying proton distribution, typically at a few to a few tens of GeV with a softening of spectral index by roughly one power from the lower to higher energy side of the break, the origin of which is still not clearly understood. Our current understanding and the major unknowns of the physics involved in particle acceleration and non-thermal emission at a radiative shock is covered by a review by \citet{Bykov2013}.

Interestingly, the shocks of middle-aged SNRs are usually found to be propagating through high-density environments such as adjacent massive molecular clouds, evidenced by observational features like 1720~MHz OH masers and various optical and infrared emission lines. Signatures like strong forbidden line emission around the shocks suggest that the shocks have undergone transition into the radiative phase.\footnote{A broader overview on the rich observational properties of these objects can be found, e.g., in the recent review article by \citet{SlaneEtal2014}.} It is thus natural to question whether radiative shocks in molecular clouds can somehow manage to generate bright non-thermal radio and $\gamma$-ray emission with the observed properties, despite their low velocities compared to non-radiative shocks at younger SNRs. Several theoretical models have already been presented in the literature even before the maturity of $\gamma$-ray astronomy. For example, \citet{Laan1962} showed that compression of cosmic ray electrons and magnetic field in an isothermal shock could account for the radio synchrotron emission of older shell-type supernova remnants such as the Cygnus Loop.  \citet{BC82} extended this work, and \citet{Cox1999} applied the idea to W44.  With the advent of $\gamma$-ray observations, \citet{uchiyamaea10} (hereafter `U10') advanced this scenario and constructed a phenomenological model using an analytical approach. They concluded that fast radiative J-shocks are able to pick up and re-accelerate pre-existing CRs. If these re-accelerated CRs are subsequently boosted further to higher energy density by compression inside a rapidly cooling and contracting shell behind the radiative shock, they can simultaneously produce sufficiently bright GeV $\gamma$-ray and radio synchrotron emission to explain observations. Most recently, \citet{Tang2014b} attempted to construct a time-dependent description of DSA at slow shocks based on a simple analytic hydrodynamic model for shock-cloud interactions \citep{Tang2014a} and a parametric model for the CR diffusion coefficient, and suggested that a time-dependent test-particle solution of re-acceleration under a Kraichnan-like diffusion coefficient ($D(p) \propto p^{0.5}$) can explain the observed $\gamma$-ray spectra of SNR IC~443 and W44. 

In this paper, we investigate the immediate locality of one of these cloud shocks and explore its broadband non-thermal emission mechanisms using a fully time-dependent hydrodynamical simulation, self-consistently coupled to an explicit treatment of DSA (either re-acceleration of pre-existing CRs or direct acceleration of the downstream thermal particles) and other important physical processes including those associated with a radiative shock. We conclude that, despite minor differences, our results broadly confirm the assertions by U10 that a re-acceleration model can well explain the observed broadband emission properties. A direct acceleration model on the other hand suffers from several difficulties that we will explain in detail. In Section~\ref{model}, we introduce the essential physical components of our model and explain their relevance to the problem. Major assumptions made are also stated and elaborated. We then present our results on the hydrodynamical evolution and non-thermal emission calculation in Section~\ref{results}, followed by discussions and conclusion in Section~\ref{summary}.

\section{Model}
\label{model}

In this section, we review the physics and numerical setup of our model. 
The numerical calculations are performed using the \textit{CR-hydro-NEI} simulation code \citep[][]{LEN2012}. 
In our model, we consider a picture in which the progenitor star undergoes core-collapse and explodes into a tenuous wind cavity surrounded by a molecular cloud \citep[e.g.][]{Koo1995}. The blastwave created by the expanding ejecta propagates into the cavity with high speed and eventually hits the interface with the surrounding dense medium (probably part of the parent molecular cloud of the progenitor star), typically at a radius $\sim 10$~pc from the explosion center \citep{Chevalier99}. It then penetrates into the cloud (we will call it a `cloud shock' hereafter) and keeps pushing through the dense medium, and eventually becomes radiative.
The molecular cloud is approximated as a uniform dense medium with a number density $n_0 \sim 100$~cm$^{-3}$. 
While the real situation is expected to be more complex with three-dimensional details such as clumpiness (i.e., density inhomogeneity) of the molecular cloud \citep[e.g.,][]{Rho1994}, we believe our spherically symmetric model can capture the essential physical phenomena and provide a fundamental understanding of such systems. This step is an essential precursor to more complicated and time-consuming three-dimensional hydrodynamical or MHD simulations that self-consistently include DSA and other important physical processes. 

We will trace the time evolution of the hydrodynamics of the cloud shock starting from the time of penetration, passing the point when the shock has decelerated to about 200~km s$^{-1}$ and become fully radiative. Physics including particle acceleration through the diffusive shock acceleration (DSA) mechanism and microphysical processes such as ionization and recombination, thermal conduction, radiative cooling and photoionization heating are calculated simultaneously. At the same time, charge exchange can be an important effect in astrophysical shocks. Concerning the physical conditions we are investigating here, the balance between ionization and recombination can be very sensitive to temperature and density, with additional complications due to heating terms such as photoionization. Unfortunately, our model does not currently include an accurate treatment for the effects of charge exchange. We recognize that charge exchange could have an important effect on our results, and we will address those effects in a subsequent paper. This work also does not concern the reverse shock and the ejecta emission which are relevant to the thermal X-ray emission. Such discussions are reserved to another line of studies we are carrying out in parallel, such as \citet{LPENS2014} and \citet{PL2015}, which will be merged with the current work in the future for a more complete description of middle-aged SNRs.

The typical timescale at which the shock becomes radiative for a supernova (SN) explosion of energy $E_{51} \equiv E_\mrm{SN}/(10^{51}$~erg) inside a uniform surrounding medium of density $n_0$ is given by \citet{Blondin1998}:
\begin{equation}
t_\mathrm{tr} \approx 2.9 \times 10^4 E_{51}^{4/17} n_0^{-9/17}\ \mathrm{yr}\ ,
\label{t_tr}
\end{equation}
By replacing $E_\mrm{SN}$ with the total energy of the driving blastwave in the cavity and $n_0$ with the cloud density, this quantity is found to be a convenient time unit for measuring the radiative transition of the transmitted cloud shock as well, provided that effects of efficient particle acceleration and other sources of energy loss which can further speed up the deceleration of the cloud shock are not significant. Hence, we use it as the basic time unit for our models throughout this work. Rapid cooling occurs at about $t \gtrsim t_\mathrm{tr}$ at which a cold, condensed dense shell should form behind the radiative shock where it is pushed outward by the hotter interior. Compression of the gas and the transverse component of the $B$-field lines is expected inside the contracting cool shell. The same should also occur with the trapped (re-)accelerated CR particles inside the shell which not only increase in density but are also energized by the compression, i.e., 
\begin{equation}
\Delta p = (s^{1/3}-1)p ~,
\end{equation}
where $\Delta p$ is the momentum increase of a CR particle experiencing the compression and $s \gg 1$ is the compression ratio in the cool shell.
We assume that the $B$-field is frozen into the plasma behind the shock. 

The enthalpy density $H$ in a Lagrangian gas element can be written as:
\begin{equation}
H = \frac{\gamma_g}{\gamma_g-1}P_\mrm{th} + \frac{\gamma_\mrm{CR}}{\gamma_\mrm{CR}-1}P_\mrm{CR} + \frac{B_\perp^2}{4\pi} + \frac{1}{2}\rho v^2\ , 
\end{equation}
where $\rho$ is the mass density, $v$ is the flow speed, $B_\perp$ is the $B$-field component perpendicular to the shock normal, $P_\mrm{th}$ and $P_\mrm{CR}$ are the thermal and CR pressures, $\gamma_g$ and $\gamma_\mrm{CR}$ are the ratio of specific heats for ideal gas and CRs respectively. The evolution of enthalpy per unit mass $h \equiv H/\rho$ in the gas element is followed using the energy equation below \citep[e.g.,][]{Cox1972},
%
\begin{equation}
\frac{dh}{dt} = \frac{-n_e n_H\Lambda(x) + n\Gamma(x) + \kappa \nabla^2 T}{\rho}\ ,
\end{equation}
where $n_e$, $n_H$ and $n$ are the electron, hydrogen and total number densities respectively, 
and $\Lambda$, $\Gamma$ and $\kappa$ are the cooling coefficient, heating coefficient and thermal conductivity respectively. 
The treatment of each of these components will be described below in more detail. 

In the immediate post-shock region, it has been suggested by observations that the electron-to-proton temperature ratio is small for shocks faster than 1000~km~s$^{-1}$, but close to 1 for slower shocks such as those we are interested in here \citep[e.g.,][]{Ghavamian2001,GLR2007,Rakowski2008,Helder2010}. We hence adopt an instantaneous post-shock equilibration of temperatures among ions and electrons. 
The elemental abundances in the dense cloud are taken from the observations and models of bright atomic lines from SNR molecular shocks in the IR band by \citet{Reach2000}.
All molecular species are expected to be fully dissociated by the fast cloud shock.  

\subsection{Non-equilibrium Ionization, Cooling and Heating}
\label{cooling}

The fully time-dependent ionization and recombination of 12 chemical elements including H, He, C, N, O, Mg, Ne, Si, S, Ar, Ca and Fe are followed behind the shock. The number densities of a total of 152 ion states are calculated in each gas element and time step in order to estimate effects like radiative cooling and photoionization heating self-consistently. The details of our treatment of non-equilibrium ionization (NEI) can be found in \citet{PES2009,PSRE2010}.

By tracing the ionization states and number densities of all ion species in the post-shock plasma through the NEI calculation, we can compute their contributions to radiative cooling through various line and continuum emissions at each time step. This approach is more accurate than applying a power-law cooling curve which only depends on the temperature. A selection of 12 relatively strong optical and infrared collisional excited lines (including [C II] 156$\mu$m, [N II] $\lambda$$\lambda$6548, 6584, [O I] $\lambda$$\lambda$6300, 6364, [O II] $\lambda$$\lambda$3727, 3729, [O III] $\lambda$$\lambda$4959, 5007, [O IV] 25.8$\mu$m, [Ne II] 12.8$\mu$m, [Si II] 35$\mu$m, [S II] $\lambda\lambda$6717, 6727 and [Fe II] 26$\mu$m) are also computed. In the model considered in this work, the cooling timescale is typically much smaller than the dynamical timescale (sound crossing time) in the post-shock region of rapid recombination and radiative cooling, hence the cooling there can be approximated essentially as an isochoric process. This region is then subsequently pushed upon and compressed by the faster and hotter interior to form a dense shell until a mechanical equilibrium is approached. This equilibrium in the dense cool shell is predominantly supported by the non-thermal (i.e., magnetic and cosmic-ray) pressures, as will be discussed further later in this paper. During the transition to the pressure-driven phase, large amplitude oscillations of the radiative cloud shock due to thermal instability are expected in an over-stable fashion until the shock velocity becomes lower than $\sim 120$~km~s$^{-1}$ \citep[e.g.,][]{CI1982, Kimoto1997, Blondin1998}, as we will also clearly show below.
We do not consider cooling and heating effects from molecular chemistry like molecule reformation and IR line emission below a few 100~K.   

Absorption of and photoionization by locally generated UV photons can heat up the gas by the production of photoelectrons. For simplicity, we assume a temperature of $T_0 \sim 10^4$~K in the precursor due to photoionization heating after the shock becomes radiative. We also do not treat the time-dependent radiative transfer in the post-shock gas. We calculate the heating rate $\Gamma(x)$ by estimating the fluxes and optical depths of photoionizing photons including the continuum emission and important emission lines, such as He I, He II, O III-V etc, in each gas parcel. The absorption cross sections are calculated using the fitting recipe of \citet{Verner96}. Following a method similar to that in \citet{Gnat2009}, we calculate in a steady-state limit the absorption of these self-generated photons by each chemical element in the gas cells further downstream and the resulting production of photoelectrons whose residual energies heat up the plasma. 


Thermal conduction can be important at places with large spatial temperature gradients, such as the rapidly cooling gas behind a radiative shock propagating in a molecular cloud. We treat thermal conduction by parameterizing the conductivity relative to the Spitzer conductivity $\kappa_\mathrm{Spitzer}$ as
\begin{equation}
\kappa = f_\mathrm{cond} \ \kappa_\mathrm{Spitzer} = f_\mathrm{cond}  \left(1.84 \times 10^{-5} \times \frac{T^{5/2}}{\mathrm{ln}\Lambda_c}\right)~,
\end{equation}
where $\mathrm{ln}\Lambda_c \approx 37$ is the Coulomb logarithm.
In our model, we set the parameter $f_\mathrm{cond} = 0.3$ for a collisionless plasma \citep[see e.g.][]{Narayan2001,Bale2013}, but our results are not sensitive to a moderate change to this parameter.

\subsection{Non-thermal Emission from a Cloud Shock}
Currently, there still remain controversies on the true origin of the center-filled thermal X-ray emission from middle-aged, mixed-morphology SNRs. For example, it is unclear what leads to their center-filled morphology \citep[e.g.,][]{White1991,Cui1992}, and whether they are dominated by the shocked ambient matter \citep[e.g.,][]{Zhou2014} or by the ejecta \citep[e.g.,][]{Uchida2012} which most probably have to be discussed on a remnant-by-remnant basis; also, there is ongoing debate on the physical explanation for the apparent ``over-ionized'' plasma state \citep[e.g.,][]{Itoh1989,Kawasaki2002,Zhou2011,Shimizu2012}, evidenced by the detection of radiative recombination continuum components and accompanying anomalous ratios of H-like and He-like ionization states \citep[e.g.,][]{Kawasaki2002,Ozawa2009,Yamaguchi2009}. The situation for the non-thermal emission is arguably more clearcut. The major contributions to the bright radio continuum and $\gamma$-ray emissions, both presumably coming from the shell \citep[e.g.][]{AbdoEtalW442010}, are synchrotron radiation emitted by primary (i.e., direct acceleration of thermal $e^-$ or re-acceleration of pre-existing CR $e^-$ and $e^+$) and secondary (i.e., $e^-$ and $e^+$ from the decay of changed pions produced by hadronic interactions) leptons that gyrate around the post-shock $B$-field, and energetic $\gamma$-ray photons from the decay of neutral pions produced by the interaction of accelerated protons/ions with the dense gas around the shock, respectively. We will adopt this general picture in our models for the non-thermal emission.

\subsection{Particle acceleration and re-acceleration}
There are two possible scenarios that we can consider: (1) re-acceleration of pre-existing high-energy particles by the cloud shock; (2) efficient acceleration of particles through NLDSA via injection from the thermal pool behind the subshock.  

The re-acceleration scenario involves the shock acceleration of pre-existing high-energy particles wandering around the source, such as Galactic CRs, and escaped CRs from nearby younger SNRs. In this case, most of the pre-accelerated ``seed'' particles already have adequate momenta to avoid the difficulties of thermal injection due to Coulomb loss and long acceleration timescales. One difficulty can be the relatively low densities of the seed particles compared to the observed high luminosities of the radio and GeV emission. We will however show that this hurdle can be overcome if the shock undergoes a transition into the radiative phase. 

In this model, following U10, we assume the seed particles to be the Galactic CR protons and electrons+positrons (hereafter we refer to them simply as electrons) and adopt spectra in momentum space in the following forms,
\begin{equation}
\begin{split}
n_{p,\mrm{seed}}(p) &= 4\pi J_p \beta^{1.5} p_0^{-2.76},~p_0 > 0.31~, \\
n_{e,\mrm{seed}}(p) &= 4\pi J_e (1+p_0^2)^{-0.55} p_0^{-2},~p_0 > 0.02~,
\end{split}
\label{seed}
\end{equation} 
where $p_0 \equiv p/(\mrm{GeV}c^{-1})$, $\beta \equiv v_i/c$, $v_i$ is the particle velocity. Here $n_\mrm{seed}(p)dp$ is the particle number densities in the interval $p \sim p+dp$ which can be rewritten as $4\pi p^2 f_\mrm{seed}(p)dp$ in terms of the phase space distribution function $f_\mrm{seed}(p)$. The normalization factors adopted are $J_p = 1.9$~cm$^{-2}$s$^{-1}$sr$^{-1}$GeV$^{-1}$ and $J_e = 0.02$~cm$^{-2}$s$^{-1}$sr$^{-1}$GeV$^{-1}$, so the total energy densities of the seed protons and electrons are $\varepsilon_p \approx 0.81$~eV~cm$^{-3}$ and $\varepsilon_e \approx 4.8 \times 10^{-3}$~eV~cm$^{-3}$ respectively. With such, we can calculate the spectra of the re-accelerated particles using an iterative semi-analytic approach described in \citet[]{Blasi2004} and \citet{LEN2012}, as follows,
\begin{eqnarray}
f_i(p) = &&\frac{3S_\mathrm{tot}}{S_\mathrm{tot}U(p)-1}  \int_{p_\mathrm{min}}^p \frac{dp''}{p''} \Bigg\{ f_{i,\mrm{seed}}(p'') \nonumber \\ 
&& \times ~\mathrm{exp} \left[ -\int_{p''}^p \frac{dp'}{p'} \frac{3S_\mathrm{tot}U(p')}{S_\mathrm{tot}U(p')-1} \right]\Bigg\}  \nonumber \\
&& \times ~\mrm{exp}\left[-\left(\frac{p}{p_{\mrm{max},i}}\right)^\alpha\right]~,  
\label{eqn:f0}
\end{eqnarray}
where $i$ represents $e^{-/+}$ or $p$, $U(p)$ and $S_\mrm{tot}$ are the dimensionless gas flow velocity in the shock rest frame\footnote{$U(p)$ ($\le 1$) is normalized to the shock speed, and is a function of $x$ but with a change of variable to $p$ using the momentum-dependent diffusion length of a CR particle in the shock precursor. It represents the smoothed shock structure when CR back-pressure strongly modifies the incoming flow speed profile. For example, $U(p) \approx 1$ for all $p$ if shock modification from CR pressure is unimportant, and $U(p) < 1$ when DSA is efficient due to a deceleration of the incoming flow by the back-streaming CRs in the shock rest frame.} and the compression ratio experienced by the streaming CRs in the shock precursor respectively, $p_\mrm{min}$ is the corresponding low-energy cutoff of the seed spectra specified in Equation~(\ref{seed}), and $p_\mrm{max}$ is the maximum momentum. The index $\alpha$ specifies the rollover shape near $p_\mrm{max}$ and we adopt $\alpha = 1.5$ for our models. For detailed definitions of these quantities, see \citet{LEN2012}. We do not consider the acceleration of heavy ions or dust grains here. 

For the NLDSA scenario, we expect that DSA at the cloud shock with a velocity $v_\mrm{sk} \sim 100$~km s$^{-1}$ and density $n_0 > 100$~cm$^{-3}$ may face difficulty if the so-called `thermal leakage' mechanism is considered for the injection of suprathermal particles into the DSA process. A first possible difficulty is the fast energy loss of the injected particles through Coulomb interactions and ionization which competes against their early acceleration and hence hinders their path towards higher energies. This constraint is irrelevant if the ratio $t_\mrm{acc}/\mrm{min}(t_\mrm{Coul}, t_\mrm{ion}) \ll 1$, where $t_\mrm{acc}$, $t_\mrm{Coul}$ and $t_\mrm{ion}$ are the acceleration time-scale, Coulomb loss time-scale and ionization loss time-scale respectively at around the injection energy. For a temperature $\sim 10^4$~K and $B$-field strength $\sim 10~\mu$G, this condition can be expressed approximately as $(v_\mrm{sk}/100\mrm{~km~s}^{-1})^2/(n_0/1\mrm{~cm}^{-3}) \gg 10^{-5}$ \citep{DDK96}, which does hold for $n_0 \sim 100$~cm~s$^{-1}$ and  $v_\mrm{sk} \sim 100$~km~s$^{-1}$. Furthermore, in situations where the cloud is highly clumpy, the shock can be propagating in an inter-clump medium \citep[e.g.,][]{Tang2014a} which has a much lower average density than the cloud cores and hence poses less problem from Coulomb and ionization losses. 

However, even if Coulomb and ionization losses are not problematic, the slow shock speed still implies a long acceleration time required for boosting the particles from the injection energy up to $\gamma$-ray emitting energies, typically $\gtrsim 100$~GeV. One possibility to overcome this is to consider a situation with significant amplification of the magnetic turbulence, such as by CR-driven instabilities \citep[e.g.,][]{Bykov2014}, to shorten the acceleration time. In this paper, we are going to discuss such a case with an efficient NLDSA of particles injected from the thermal pool, which can produce strongly amplified $B$-field through CR steaming instability in the shock precursor. Our formulation for NLDSA and its feedbacks to the hydrodynamics can be found in \citet{LEN2012} and references therein.

\subsubsection{Diffusion Coefficient and Momentum Break}

\begin{figure*}
\centering
\subfloat[]{\includegraphics[width=7.5cm]{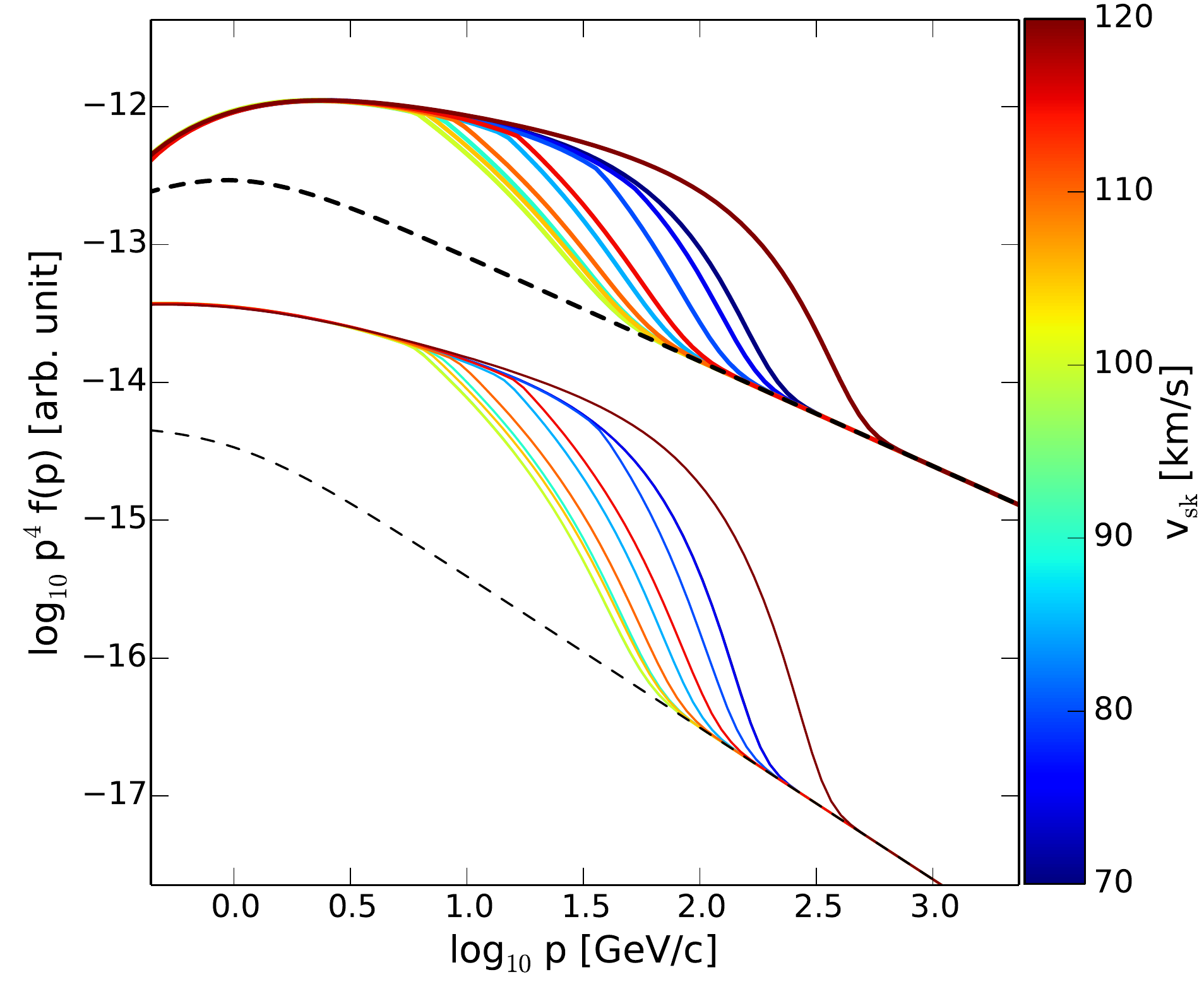}}
\subfloat[]{\includegraphics[width=7.7cm]{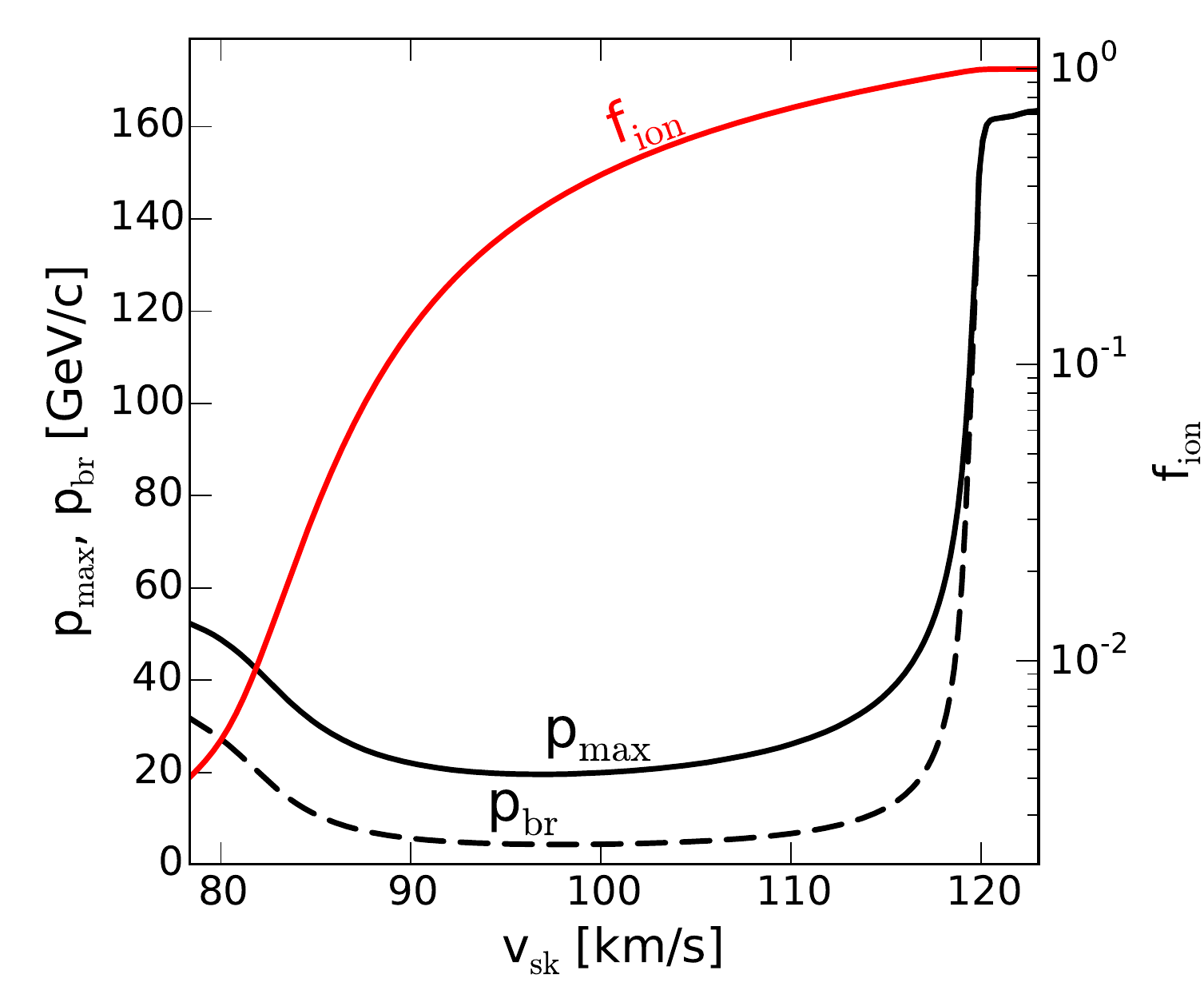}}
\caption{(a) Spectra of re-accelerated Galactic CR at a steady cloud shock of velocities $v_\mrm{sk} < 120$~km s$^{-1}$. Here, the ambient magnetic field $B_0 = 25~\mu$G and $T_0 = 10^4$~K in the shock precursor. The precursor ion fraction $f_\mrm{ion}$ is estimated according to \citet{HM89} which changes the break momentum $p_\mrm{br}$ of the re-accelerated particles. The dashed lines show the ambient spectra of the CRs before re-acceleration. The upper thick and lower thin lines correspond to protons and electrons respectively.
(b) Variation of esacpe-limited maximum momentum $p_\mrm{max}$, break momentum $p_\mrm{br}$ and precursor ion fraction $f_\mrm{ion}$ with shock velocity $v_\mrm{sk}$ in correspondence with the spectra shown in panel (a).} 
\label{fp}
\end{figure*}
It is straightforward to confirm that a Kolmogorov or Kraichnan magnetic turbulence spectrum typically associated with the interstellar medium (ISM) or the insides of giant molecular clouds ($D \sim10^{27}$ to $10^{28}$~cm$^2$~s$^{-1}$ at 10~GeVc$^{-1}$) is unable to support acceleration of particles to energies that can generate substantial sub-TeV photons at a slow cloud shock in a dense medium. It has been shown that stronger turbulence which reduces the diffusion coefficient by a factor of $\sim 100$ relative to the Galactic mean is necessary to reproduce the observed $\gamma$-ray fluxes of middle-aged SNRs \citep[e.g.,][]{FujitaEtal2009,OhiraEtal2011,Yan2012}. 
Observationally, there is evidence for such turbulent shock precursors from recent H$\alpha$ observations of SNR shocks interacting with gas clouds \citep[e.g.,][]{JJ2010,Medina2014}.

One possibility for the presence of strong magnetic turbulence in the molecular clouds is an efficient, non-linear CR acceleration through DSA at the cloud shock which can amplify the turbulent component of the upstream $B$-field through various CR-driven instabilities \citep[e.g.,][]{Bykov2014}, which applies to our NLDSA scenario. 
Another possibility is that efficient acceleration occurred in the past when the SNR shock was propagating at high velocities in the wind cavity, from which high-energy CRs that escaped upstream from the shock can penetrate through the surrounding cloud and generate strong magnetic turbulence there through the streaming instability \citep{Yan2012}. 

In either scenario, we assume that the $B$-field in the molecular cloud is highly turbulent and fully tangled to support Bohm-like diffusion of CRs up to a certain particle momentum above which the ion-neutral damping effect in a partially neutral medium takes place. At the cloud shock, the upstream medium can be expected to be partially ionized. Even after the shock has become radiative and the UV light produced in the downstream can create a photoionization precursor, the ionization is partial and neutral particles exist for $v_\mrm{sk}\ \lsim 120$~km s$^{-1}$. In such a case, ion-neutral damping of the magnetic turbulence in the shock precursor will occur and hamper CR acceleration at the highest momenta. More specifically, the pitch-angle scatterings of the CRs with the turbulent $B$-field will experience a transition at a certain momentum $p_\mathrm{br}$ above which wave-particle scatterings become weaker and less efficient.

At around $p_\mathrm{br}$, the CR diffusion in the precursor changes from Bohm to a regime in which the CRs scatter only with small-scale waves and can be described by a diffusion coefficient $D(p) \propto p^2$. In this light, once the shock velocity has dropped to $v_\mrm{sk}\ \le 120$~km s$^{-1}$, we adopt a diffusion coefficient for the CRs with a smooth break at $p_\mathrm{br}$ with the following form:
\begin{equation}
D(x,p) = \eta_B D_B(x,p)\left(1+\frac{p}{p_\mathrm{br}}\right)
\label{diff_coef}
\end{equation} 
where $D_B$ is the Bohm diffusion coefficient and $\eta_B \sim 1$. 
For a radiative shock with higher velocities, full ionization is assumed in the precursor and $D(x,p) = \eta_B D_B(x,p)$ is used. 

We can estimate the break momentum by following \citet{MDS2011} and references therein, i.e.,
\begin{equation}
p_\mathrm{br} = 10 \left(\frac{T_0}{10^4 \mathrm{K}}\right)^{-0.4} \left(\frac{B_0}{\mu \mathrm{G}}\right)^2 n_n^{-1} n_i^{-1/2} m_p c~,
\label{pbr}
\end{equation}
where $n_n$ and $n_i$ are the neutral and ion densities. The ion fraction $f_\mrm{ion} \equiv n_i/(n_i+n_n)$ in the precursor of the cloud shock can be estimated approximately from $v_\mrm{sk}$ using the results of \citet{HM89}. A spectral break for the CR spectra can be expected if $p_\mrm{br} < p_\mrm{max}$, and we can impose the break by multiplying a factor of $(p/p_\mrm{br})^{-1}$ to $f_i(p)$ for $p \ge p_\mrm{br}$. \citet{Draine1993} and \citet{DDK96} also estimated the maximum energy attainable by the accelerating CRs when ion-neutral damping limits DSA. The maximum energies they derived have the same ion fraction dependence as $p_\mrm{br}$ in Equation~\ref{pbr}.     

The ion fraction $f_\mrm{ion}$ is estimated approximately using the steady-flow equilibrium value from \citet{HM89} in our model. We note however that non-steady-state effects can bring about important modifications. For example, \citet{Cox1972} has pointed out that an enormous burst of EUV photons when the cool shell forms can photoionize the precursor, and the recombination timescale is longer than the time for further slowing of the shell. This can lead to a more highly ionized gas compared to the equilibrium values at $v_\mrm{sk} < 120$~km~s$^{-1}$. Also, $f_\mrm{ion}$ can change if non-linear feedback of DSA to the shock structure is considered \citep{Bykov2013}, and the formulae in \citet{HM89} must be modified; shock modification by the CR pressure results in an increase of the total compression ratio and decrease of the subshock compression ratio, which leads to a reduction of the post-shock temperature relative to the standard value for unmodified shocks. This hastens the transition of the shock to the radiative phase, and the ion fraction can start to increase due to photoionization at a higher shock velocity. In our models, however, DSA does not enter the nonlinear regime for the scenario of Galactic CR re-acceleration. And for the case of NLDSA of thermally injected particles, the evolution of the ion fraction in the precursor is expected to be highly non-trivial, especially when the \textit{time-dependent} NLDSA feedback to the shock hydrodynamics is accounted for. In this work, we neglect this nonlinear effect on the ion fraction for simplicity.   

\subsubsection{Spectrum of Accelerated Particles}

In the case of re-acceleration, the maximum momentum $p_\mrm{max}$ of the CR protons at the cloud shock is most likely determined by their escape far upstream. We model their escape through setting a free-escape-boundary (FEB) at 10\% of the current shock radius ahead of the shock, so that CRs with diffusion lengths longer than the FEB in the precursor are considered to have escaped.

For particle acceleration via injection from the thermal pool, $p_\mrm{max}$ is additionally limited by the finite acceleration time due to the slow shock speed in a molecular cloud; in fact, significant magnetic field amplification through CR-driven instabilities in the precursor is necessary to shorten the acceleration time so that the injected thermal particles can reach sub-TeV energies in a reasonable time, therefore efficient NLDSA is required.

When ion-neutral damping in the shock precursor is important, either before the shock becomes radiative or as the radiative shock has decelerated to a velocity $< 120$~km~s$^{-1}$,  $p_\mrm{max}$ is limited due to the transition of $D(x,p)$ to a fast regime above the break momentum (i.e., waves with frequencies $\omega=kv_A$ smaller than the ion-neutral collision frequency $\nu_\mrm{in}$ become strongly damped, with $k$ and $v_A$ being the wave number and Alfv\'{e}n velocity). The $p_\mrm{max}$ of the CR electrons can be further limited by energy loss processes such as synchrotron and inverse-Compton losses if important. For more details on the calculation of $p_\mrm{max}$, see the descriptions in \citet{LEN2012}.

As an example, Figure~\ref{fp} (left) shows the spectra of re-accelerated Galactic CR protons and electrons at a steady-state cloud shock with different $v_\mrm{sk}$ in a dense medium with ambient proton density $n_0 = 200$~cm$^{-3}$, magnetic field $B_0 = 25~\mu$G and temperature $T_0 = 10^4$~K. The pre-shock ion fraction $f_\mrm{ion}$ of the incoming gas is calculated according to \citet{HM89} as described above. The cloud shock radius from the explosion center is set at $10$~pc here, but the result is not sensitive to a fractional change of this value. 

Figure~\ref{fp} (right) shows the corresponding variation of $p_\mrm{max}$, $p_\mrm{br}$ and $f_\mrm{ion}$ as a function of $v_\mrm{sk}$. Here $p_\mrm{max}$ is limited by escape and is determined by equating the CR diffusion length  to the distance between the FEB and the shock front through Equation~\ref{diff_coef}. The maximum momentum of electrons (not shown) is very close to that of the protons since magnetic field amplification due to self-generation of magnetic turbulence by the CR streaming instability in the shock precursor is inefficient in the case of pure re-acceleration of Galactic CRs, so that energy loss of the re-accelerating electrons via synchrotron radiation is slow in the immediate downstream region. However, synchrotron losses can be much faster further downstream when a highly compressed $B$-field is present in the radiatively cooled dense shell formed behind the cloud shock. 

In this model, the precursor of a radiative shock with $v_\mrm{sk} \ge 120$~km s$^{-1}$ approaches full ionization by photoionization, and wave damping through ion-neutral collisions is ineffective, so no break in the diffusion coefficient and the CR spectra appears. At lowest velocities, $f_\mrm{ion} \ll 1$ and the ion-neutral collision frequency $\nu_\mrm{in}$ in the precursor is small, hence wave damping becomes less efficient and the break momentum $p_\mrm{br}$ can rise again according to Equation~\ref{pbr}. Therefore, $p_\mrm{max}$ does not scale trivially as a simple power of $v_\mrm{sk}$ in this velocity regime. For parameters adopted in this example, a minimum for $p_\mrm{br}$ is reached at $v_\mrm{sk} \approx 100$~km s$^{-1}$ where there is a comparable concentration of ions and neutrals in the pre-shock gas. However, we note that this semi-analytic result is obtained by considering steady shocks only, and can be modified if detailed wave-particle and particle-particle interactions are self-consistently followed for a time-evolving radiative shock, which is beyond the scope of this paper.

\begin{figure}
\centering
\subfloat[Time Snapshots of Hydrodynamic Profiles]{\includegraphics[width=7.5cm]{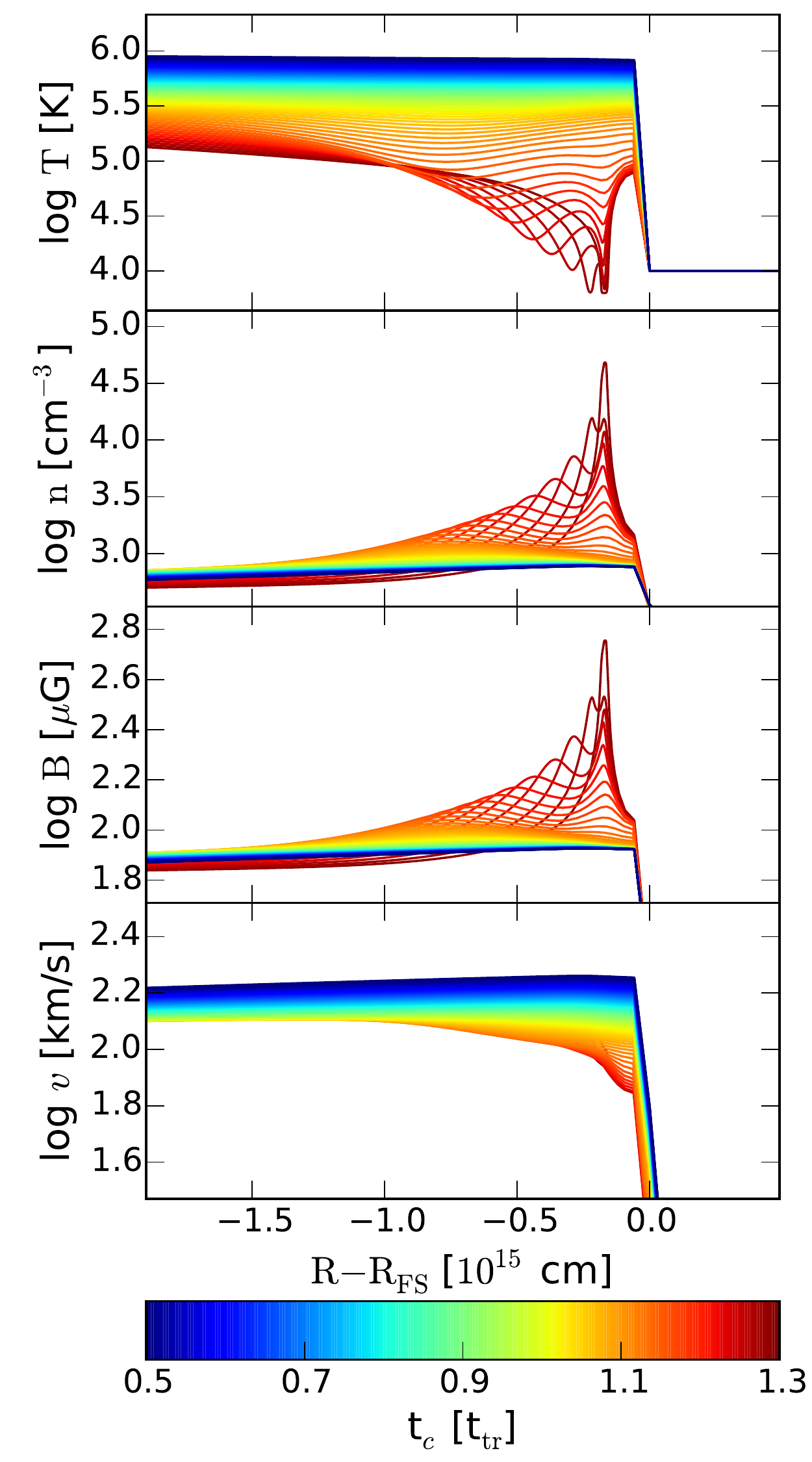}}
\newline
\subfloat[Evolution of Cloud Shock Velocity]{\includegraphics[width=6.5cm]{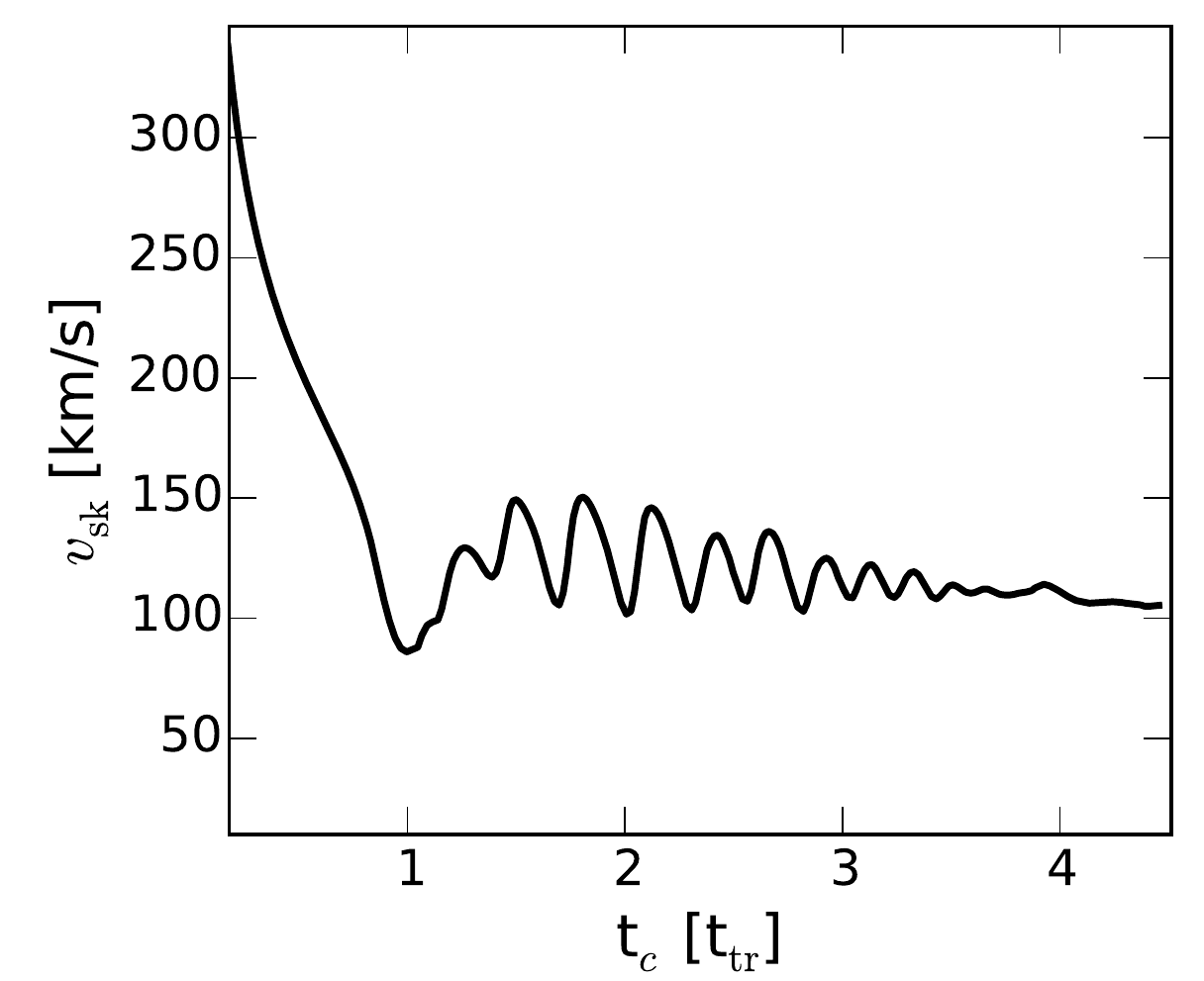}}
\caption{(a) Time snapshots of profiles of hydrodynamic variables up to the stage when the shock has just become radiative for the re-acceleration model. From top to bottom: gas temperature, gas (proton) density, $B$-field strength and gas velocity. Radius is in units of the FS radius. The colorbar depicts the evolution age of the cloud shock in units of characteristic transition time $t_\mrm{tr}$. (b) Cloud shock velocity $v_\mrm{sk}$ as a function of time for the re-acceleration scenario. 
} 
\label{hydro_profile}
\end{figure}  

\subsection{Shock Hydrodynamics in a Dense Medium}

Using the re-acceleration model as an example, we can first look at the general hydrodynamical behavior of a cloud shock just before and after its transition into the radiative phase. Here and throughout this work, we adopt a distance $d_\mrm{SNR} = 2.9$~kpc to the SNR, a final SNR radius $R_\mrm{SNR} = 12.5$~pc, an ambient gas density $n_0 = 200$~cm$^{-3}$, magnetic field $B_0 = 25$~$\mu$G and pre-shock temperature $T_0 = 10^4$~K. The pre-shock density chosen is typical of the molecular shocks at SNR W44 which are most probably responsible for producing the bright radio synchrotron filaments \citep{Reach2005, Yoshiike2013}.

The evolution of the shock hydrodynamics in the dense uniform medium is followed up to a time $t_c$.\footnote{Note that this `cloud shock time' $t_c$ is counted from the penetration of the blastwave into the cloud, so it is smaller than the actual age of the SNR.} Time snapshots of the profiles of hydrodynamic variables including total gas density $n$, temperature $T$, magnetic field strength $B$ and gas velocity $v$ are shown in Figure~\ref{hydro_profile} until the shock has just started to become radiative. When the shock decelerates quickly with time in the dense medium and the post-shock gas temperature drops to around a few $10^5$~K, radiative cooling becomes important. 
An enhancement of density and $B$-field with time due to rapid compression can be observed. 
A cool dense shell forms behind the shock which is pushed forward by a faster and hotter interior heated up earlier by a stronger shock in the past. In this phase, the dynamics approaches that of a pressure-driven blastwave. The cool shell is eventually dominated by and supported against further collapse by non-thermal pressures, as is known to be the case at radiative filaments in some remnants such as Cygnus Loop \citep[][]{Raymond1988}. The formation of the dense cool shell can happen at a time somewhat later than $1~t_\mrm{tr}$ due to the inclusion of thermal conduction which redistributes heat to slow down the cooling process. After the beginning of the transition, the cloud shock shows large oscillations in velocity for a few cycles due to thermal instability (Figure~\ref{hydro_profile}, bottom panel), as mentioned in Section~\ref{cooling}. The instability subsides and the oscillations are damped after the mean shock speed becomes lower than $\sim 120$~km~s$^{-1}$. 

\section{Results}
\label{results}

In this Section, we consider a system where the SN explosion occurred in a wind cavity in the past and the blast wave eventually hit a surrounding molecular cloud. We follow the dynamics of the transmitted cloud shock and calculate the associated radio and $\gamma$-ray emission under two scenarios of (i) Galactic CR re-acceleration, and (ii) NLDSA of thermally injected particles. To see if they can reproduce the generally observed characteristics of GeV-bright middle-aged SNRs, a comparison of the models with broadband spectral data from SNR W44 is presented.

\subsection{Case of Re-acceleration}

\subsubsection{Broadband Non-thermal SED}
 
\begin{figure}
\centering
\subfloat[Evolution of Broadband Spectrum]{\includegraphics[width=8.5cm]{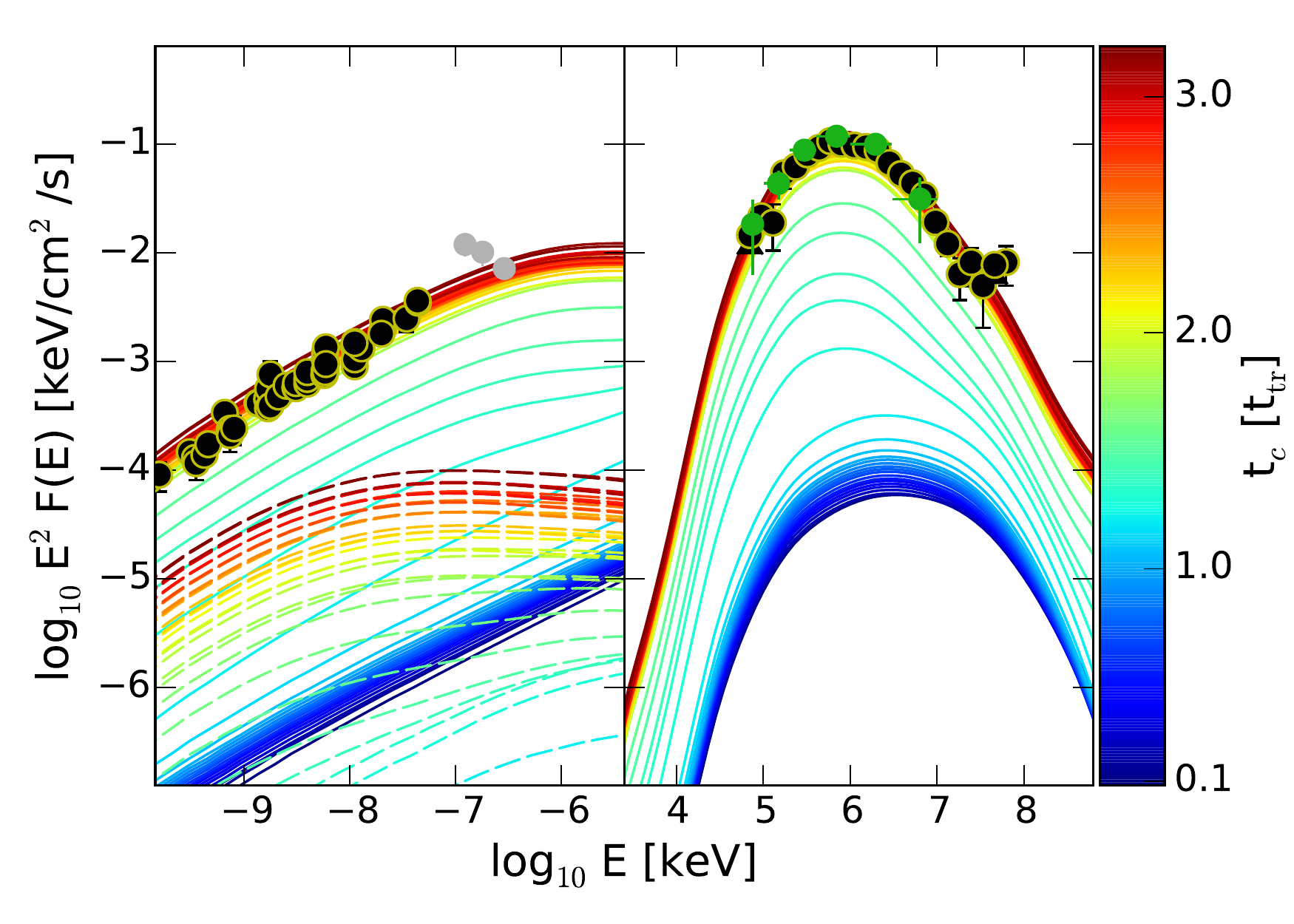}}
\newline
\subfloat[Evolution of Integrated Flux]{\includegraphics[width=8.5cm]{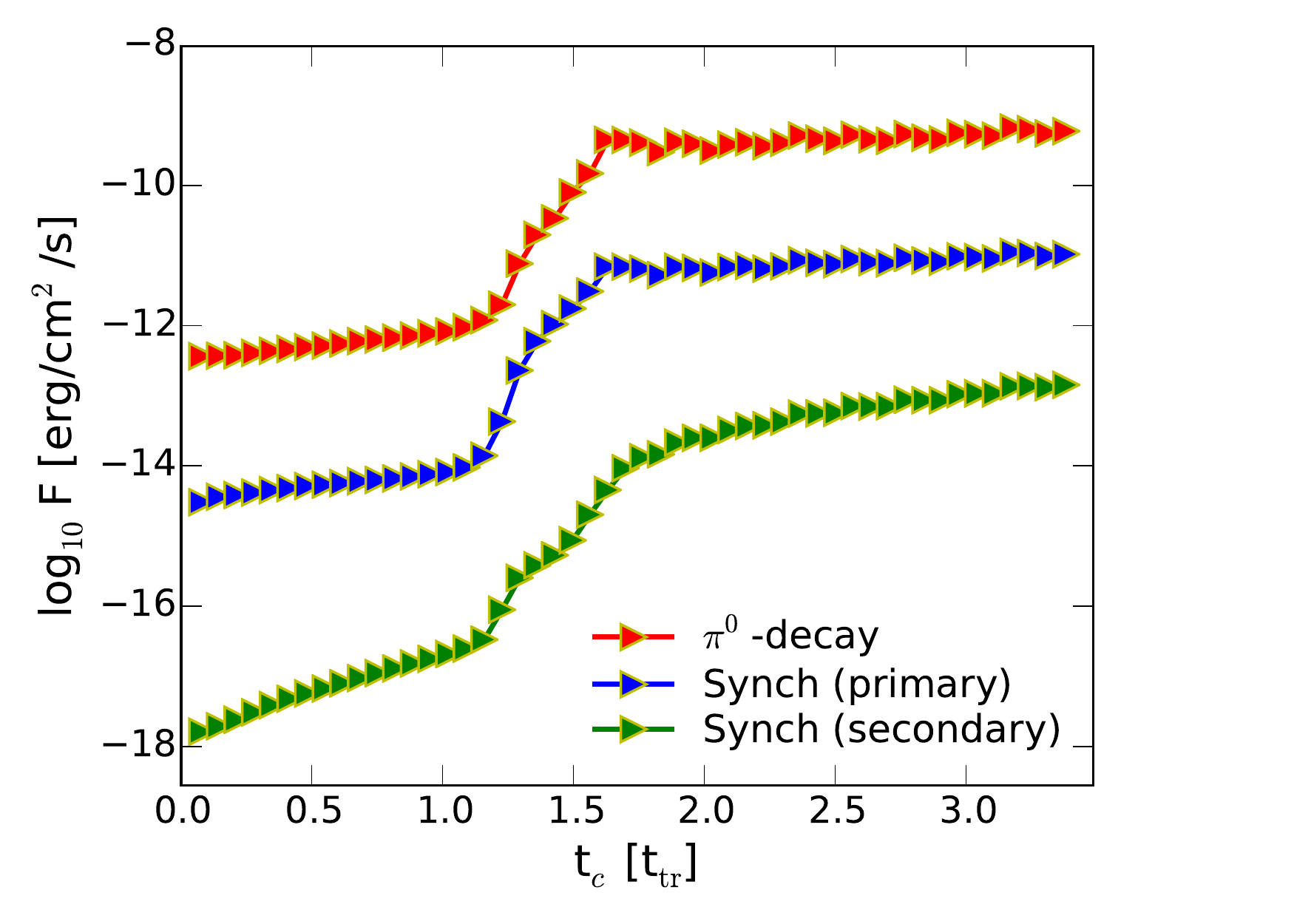}}
\caption{(a) Time evolution of the radio synchrotron (left panel) and $\pi^0$-decay $\gamma$-ray (right panel) spectra up to $t_c = 3\ t_\mrm{tr}$ of our model under the Galactic CR re-acceleration scenario. Dashed lines in the left panel show the contribution from secondary $e^-$ and $e^+$ to the synchrotron emission, while the solid lines show the total. Again, the colorbar depicts the evolution age of the cloud shock in units of $t_\mrm{tr}$. Observed data of SNR W44 from radio telescopes \citep[compilation by][black points]{Castelletti2007, Sun2011}, Planck Galactic SNR survey \citep[][grey points]{Planck2014} and $\gamma$-ray instruments including \textit{Fermi} LAT \citep[][black points]{Ackermann2013} and \textit{AGILE} \citep[][green points]{Cardillo2014} are overlaid. The radio data points are scaled by a factor of 0.5 following U10. An overall normalization factor of 0.2 is applied to the model spectra to explain the data.
\\ 
(b) The corresponding evolution of the radio synchrotron and $\pi^0$-decay $\gamma$-ray flux. The energy ranges of flux integration for the radio and $\gamma$-ray photons are 70~MeV$-$20~GeV and 70~MHz$-$10~GHz respectively.}
\label{SED_W44}
\end{figure}
Figure~\ref{SED_W44} shows the spectral evolution of the radio and $\gamma$-ray emission predicted by our model under the Galactic CR re-acceleration scenario. The corresponding evolution of integrated energy fluxes is also shown. For clarity, the contribution from non-thermal bremsstrahlung to $\gamma$-rays is not included in the plot since it is relatively unimportant compared to the $\pi^0$-decay component. The time $t_c$ is counted in units of $t_\mrm{tr}$ for generality. The average $v_\mrm{sk}$ right after the penetration is $\lsim 400$~km~s$^{-1}$.
At $t_{c} < 1.2\ t_\mrm{tr}$, both the radio continuum and $\gamma$-ray flux increase gradually with time, as more re-accelerated CRs and secondary $e^{+/-}$ from pion decay accumulate downstream. The spectral shapes do not change significantly due to the gradual evolution of the shock. Then, a prominent transition occurs at $t_c\ \approx 1.2\ t_\mrm{tr}$ when the shock has decelerated to a speed $< 200$~km~s$^{-1}$ whose dynamics start to be governed by radiative cooling, just as we have seen in Figure~\ref{hydro_profile}. The compression of the $B$-field and gas density as well as the re-energization of the re-accelerated CRs and secondary particles in the rapidly cooling and contracting gas shell boost the non-thermal fluxes both in radio and $\gamma$-ray by a factor of a few $100$. The spectral shape of the $\gamma$-rays also changes as the shock becomes slower than 120~km~s$^{-1}$ and experiences a partially ionized precursor so that the particle acceleration is hampered by the ion-neutral damping effect we discussed above. 

Eventually, as the gas cools down and compresses, the non-thermal pressure becomes higher than the thermal pressure in the shell, and the cool shell is supported by the non-thermal pressure and stops contracting. In this model, the non-thermal pressure is dominated by the (transverse) $B$-field with $P_\mrm{CR}/P_{B_\perp} \gtrsim 0.01$. Compression of the gas, CRs and $B$-field then halts in the shell. The pressure ratio $P_\mrm{th}/P_B$ (plasma-$\beta$) before and after the collapse of the cold shell for this model is $\lsim 100$ and $\lsim 1$ respectively inside the shell. Since the plasma-$\beta$ is proportional to $nT/B_\perp^2$ and $B_\perp$ reacts to the collapse roughly as $\sqrt{n}$, its evolution depends mainly on the temperature change only. Since radiative losses of CRs in the interesting energy bands are also slow compared to the dynamical timescale,  the non-thermal emission fluxes which mainly originate from the dense shell become stable after the transition. We note however that we do not consider the escape of CRs from the dense shell in our model which can result in a decay of the non-thermal fluxes with time. However, the emission here are dominated by protons and electrons in the GeV energy range embedded in a compressed $B_\perp$ of a few $100~\mu$G whose Larmor radii should be very small, and hence we may expect them to diffuse away from the shell only slowly as well. On the other hand, the continued production and accumulation of secondary $e^+e^-$ from hadronic interactions by the trapped CR protons in the dense shell gives rise to a gradual increase of the secondary synchrotron emission after the transition. The total fraction of the initial cloud shock energy attributed to CR re-acceleration is $\lsim1$\% at the end of simulation in this model. 

In the same figure we have overlaid onto our model the up-to-date observed spectra and fluxes of SNR W44 from radio telescopes \citep{Castelletti2007, Sun2011} and $\gamma$-ray instruments including \textit{Fermi} LAT \citep{Ackermann2013} and \textit{AGILE} \citep{Cardillo2014}. We find an overall satisfactory agreement of the broadband spectral shape. From \citet{Castelletti2007}, the radio spectral index $\alpha = -0.37 \pm 0.02$, and the range of index from our model after the flux has reached the maximum at $t_c > 2\ t_\mrm{tr}$ is $\alpha \approx -0.38$ to $-0.40$, which are reasonably consistent with each other or at most only slightly softer for the model. The predicted radio spectrum shows a spectral softening above $\sim10$~GHz due to synchrotron loss, and is consistent with the 70~GHz data from Planck observation \citep{Planck2014}, but it cannot explain the 30 and 44~GHz points which do not seem to agree with a simple extrapolation from the lower-frequency data. This apparent `excess' can possibly arise from the anomalous microwave emission from small spinning dust grains \citep{Scaife2007} since W44 lies at a low Galactic latitude. In this model, most of the bright $\pi^0$-decay $\gamma$-rays and radio synchrotron emission originate from the cool dense shell which is a thin compressed region behind the radiative cloud shock. This is qualitatively consistent with the filamentary morphology of the shell-like remnant revealed in the radio waveband. And as a result, the normalization of the non-thermal spectra for this model mainly depends on the compression of the cold dense shell, rather than the acceleration efficiency. To conform with the observed flux level, an overall normalization factor of 0.2 is applied to the model, which can be interpreted as a reasonable filling factor of the $\gamma$-ray and radio continuum emitting region over the whole $4\pi$ shell. 

It is interesting to note that, in spite of a similar basic scenario, a difference between our model and the analytical results of U10 can be found in the ratio between the secondary and primary synchrotron flux in the radio waveband. They predict a significant contribution of the secondary component in the longer wavelength regime. Our model predicts a much less prominent secondary component, and the primary emission alone can explain the observed spectrum reasonably well. We believe this difference stems from the following. U10 assume the presence of a non-evolving, time-independent cool dense shell formed behind a radiative shock which adiabatically compresses all injected primary particles once they are re-accelerated at the shock. Re-accelerated protons are continuously injected into such a cool dense shell and stay there to produce secondary $e^{+/-}$ at a fast rate ($t_{pp} \propto n_\mrm{gas}^{-1}$) up to the present day. Our model, on the other hand, treats the evolution of both the primary and secondary $e^{+/-}$ populations behind the shock fully coupled to the hydrodynamics, including the transition of the cloud shock into its radiative stage with a time-dependent formation of the cool dense shell, which certainly has an important impact on the production and accumulation of the secondary leptons through post-shock hadronic interactions.  

\subsubsection{`Forbidden' Line Emission}

\begin{figure}
\centering
\includegraphics[width=7cm]{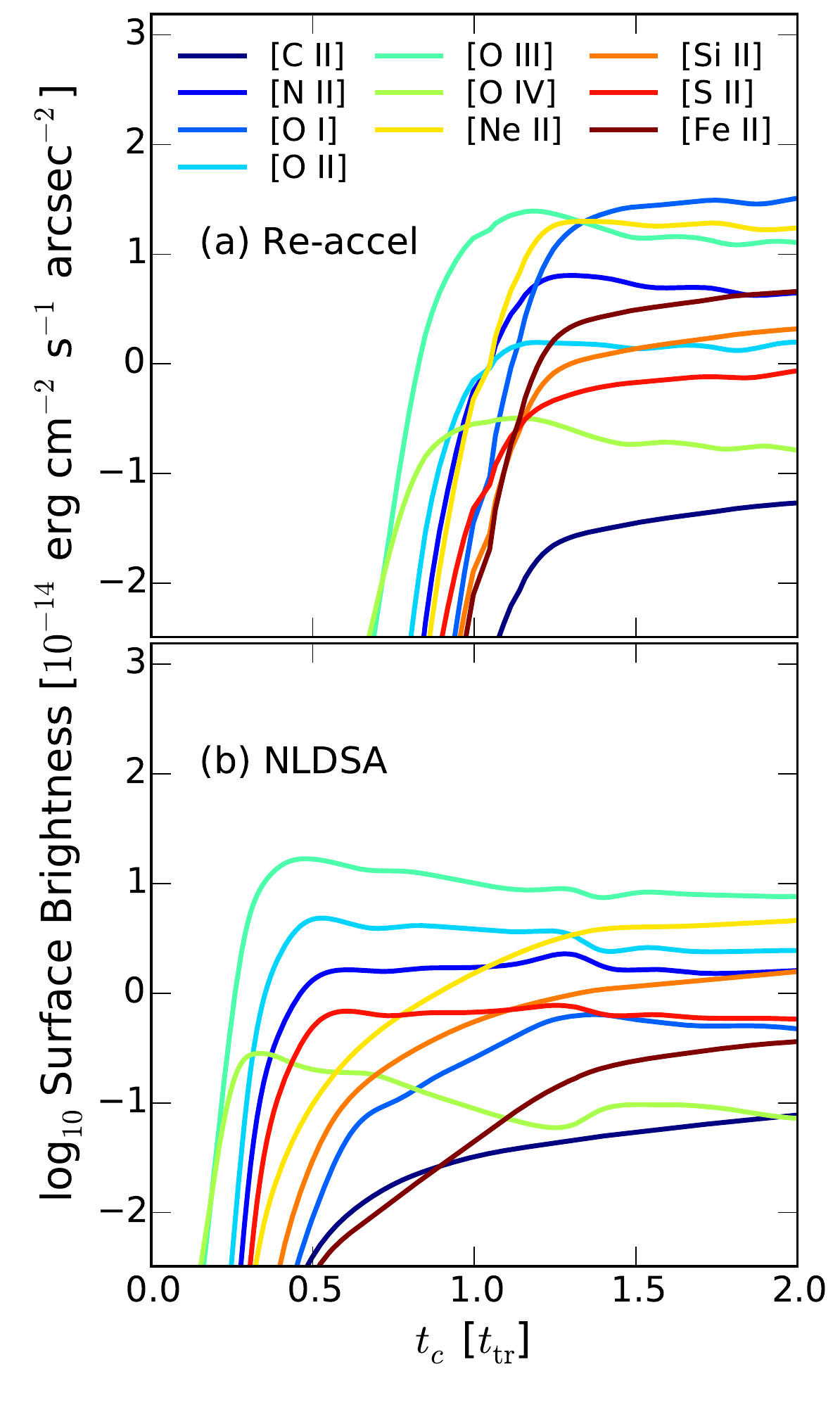}
\caption{Brightening of forbidden lines behind the cloud shock for the re-acceleration (top panel) and NLDSA model (bottom panel). For each line, the emissivities are averaged over the volume of the shocked cloud material at each time snapshot. The lines are the weighted total of their doublets, if applicable. 
}
\label{Emis_forb}
\end{figure}

Another interesting order-of-magnitude check we can perform is to 
compare the predicted emissivities of `forbidden' optical and IR collisionally excited lines behind the radiative
shock to observation. Figure~\ref{Emis_forb} shows the time evolution
of the surface brightnesses of relatively luminous lines included
in our calculation of the cooling function, up to $t_c = 2.0$ when the 
computed radio and $\gamma$-ray flux broadly agree with observations.
For the re-acceleration model, like the $\gamma$-ray and radio emission, the calculated forbidden line emission peak spatially within a thin region slightly behind the cloud shock either inside the cold dense shell or the adjacent rapidly cooling layer, resulting in a filamentary morphology following the radiative shock front. 

The re-acceleration model predicts moderately strong [\ion{O}{1}] $\lambda\lambda$(6303+6363) and [\ion{O}{3}] $\lambda\lambda$(4959+5007)
emission, but observations of W44 in [\ion{O}{1}] are not readily available
in the literature (although very bright 63$\mu$m [O I] line emission at the NE rim has been reported by \citet{Reach1996}), and the SNR is highly absorbed, so [\ion{O}{3}] emission is unlikely to be detected. 

\citet{MBG03} measured an average observed surface brightness F$_\mrm{[S~II]} \approx 4 \times$10$^{-17}$ erg cm$^{-2}$ s$^{-1}$ arcsec$^{-2}$. With their suggested E($B-V$) of 3.28, the attenuation factor by interstellar absorption is about 1700 at 6720 \AA, so the corrected surface brightness becomes F$_\mrm{[S~II]} \approx 7 \times$10$^{-14}$ erg cm$^{-2}$ s$^{-1}$ arcsec$^{-2}$. Our model predicts a [\ion{S}{2}] $\lambda\lambda($6736$+$6731) intrinsic surface brightness of order $\lesssim$ 10$^{-14}$ erg cm$^{-2}$ s$^{-1}$ arcsec$^{-2}$. Considering the effect of limb brightening from projection, this brightness should be further enhanced by a factor of a few or an order-of-magnitude, making it fairly consistent with observation.
We note that Mavromatakis et al. (2003) obtained electron densities below 220 $\rm cm^{-3}$ from the [S II] doublet ratio, while our models produce much higher densities because they start with pre-shock densities of 200 $\rm cm^{-3}$.  This is a problem for any shock model in that a high density is needed to account for the flux, but a low density is needed to match the [S II] doublet ratio.  High magnetic and CR pressures help somewhat, but some combination of lower pre-shock density, emission from the photoionization precursor and ionization and heating of the cold shell may be needed to match both the flux and electron density derived from [S II].

\subsection{Case of NLDSA with Thermal Injection} 

Here in the NLDSA scenario, we consider an efficient injection of particles from the thermal pool into the DSA process. It is not obvious whether such an efficient injection can occur at a cloud shock, but we suppose that it can happen here so that we can study its observational consequences. We fix the so-called injection parameter $\xi_\mrm{inj} = 3.7$, which means the injection momentum above which the downstream thermal particles are injected is $p_\mrm{inj} = \xi_\mrm{inj} p_\mrm{th,2}$,
where $p_\mrm{th,2} \equiv \sqrt{2m_pkT_2}$ and subscript `2' means values taken immediately downstream from the subshock. The electron-to-proton number ratio at relativistic energies $K_\mrm{ep}$ is set at $0.003$ to fit the radio-to-$\gamma$-ray flux ratio. Of course, when the shock has slowed down to a point where only a very low ion fraction exists in the photoionization precursor (below a few tens of km s$^{-1}$ as shown in Figure~\ref{fp}(b)) and the downstream electron temperature is too low for collisional ionization to occur at an appreciable rate, densities of protons and electrons should become too low for efficient injection to occur. In such phases, however, acceleration should be very inefficient anyway due to the very small $v_\mrm{sk}$. The introduction of a spectral break by ion-neutral collisions at low $v_\mrm{sk}$ is also effective on quenching efficient nonlinear acceleration. As pointed out by \citet{MDS2011}, $p_\mrm{br}$ essentially acts as $p_\mrm{max}$ in the conventional NLDSA picture since most of the CR partial pressure is contributed by the highest energy particles. The break hence causes a large reduction of CR back-reaction and shock compression ratio. Therefore, contributions to the non-thermal emission in this phase are relatively unimportant. Most of the high-energy particles contributing to the non-thermal emission are produced in the earlier phase in the evolution of the cloud shock, right after the blastwave penetration across the wind-cloud interface.

\begin{figure}
\centering
\subfloat[Evolution of Broadband Spectrum]{\includegraphics[width=8.5cm]{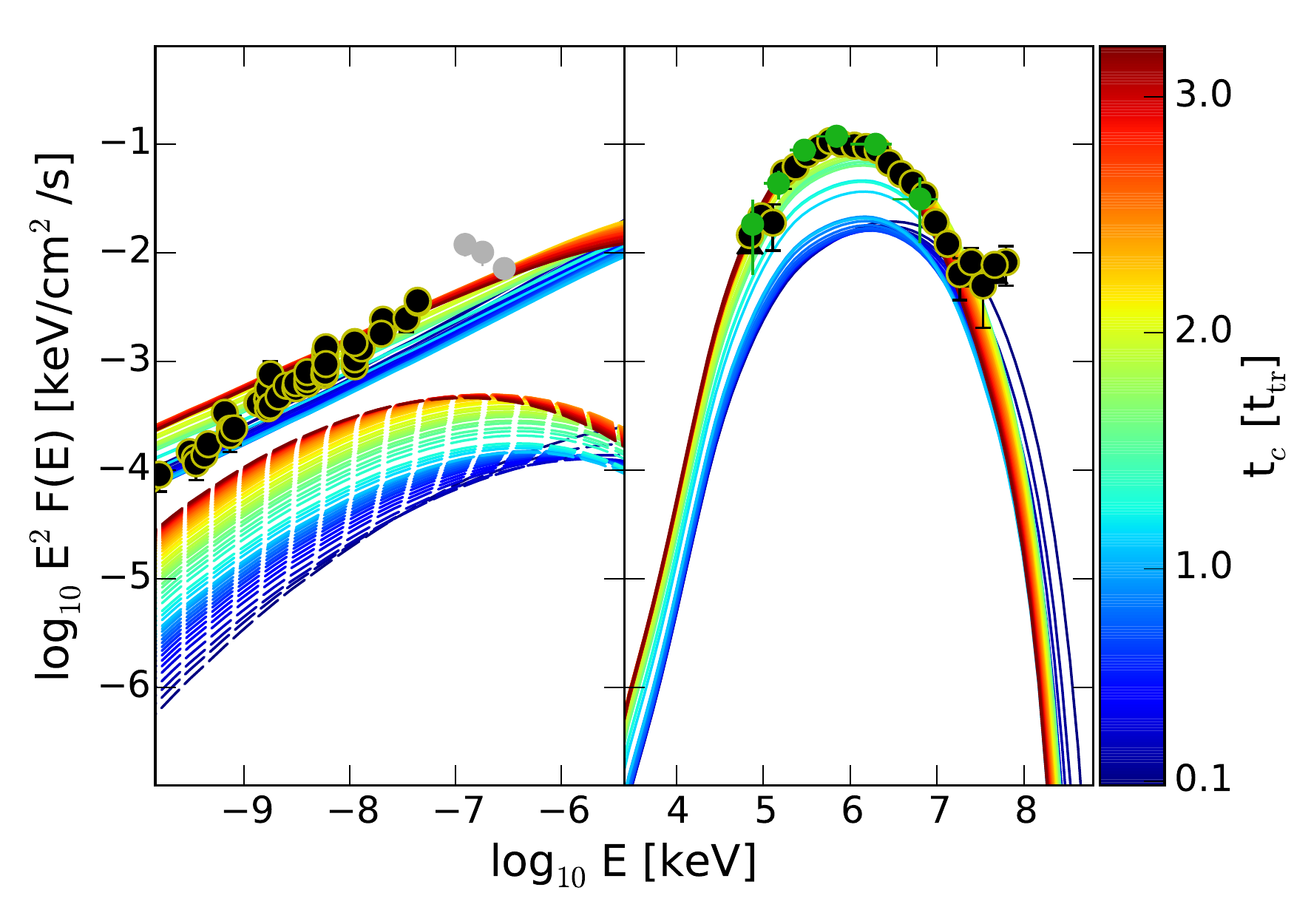}}
\newline
\subfloat[Evolution of Integrated Flux]{\includegraphics[width=8.5cm]{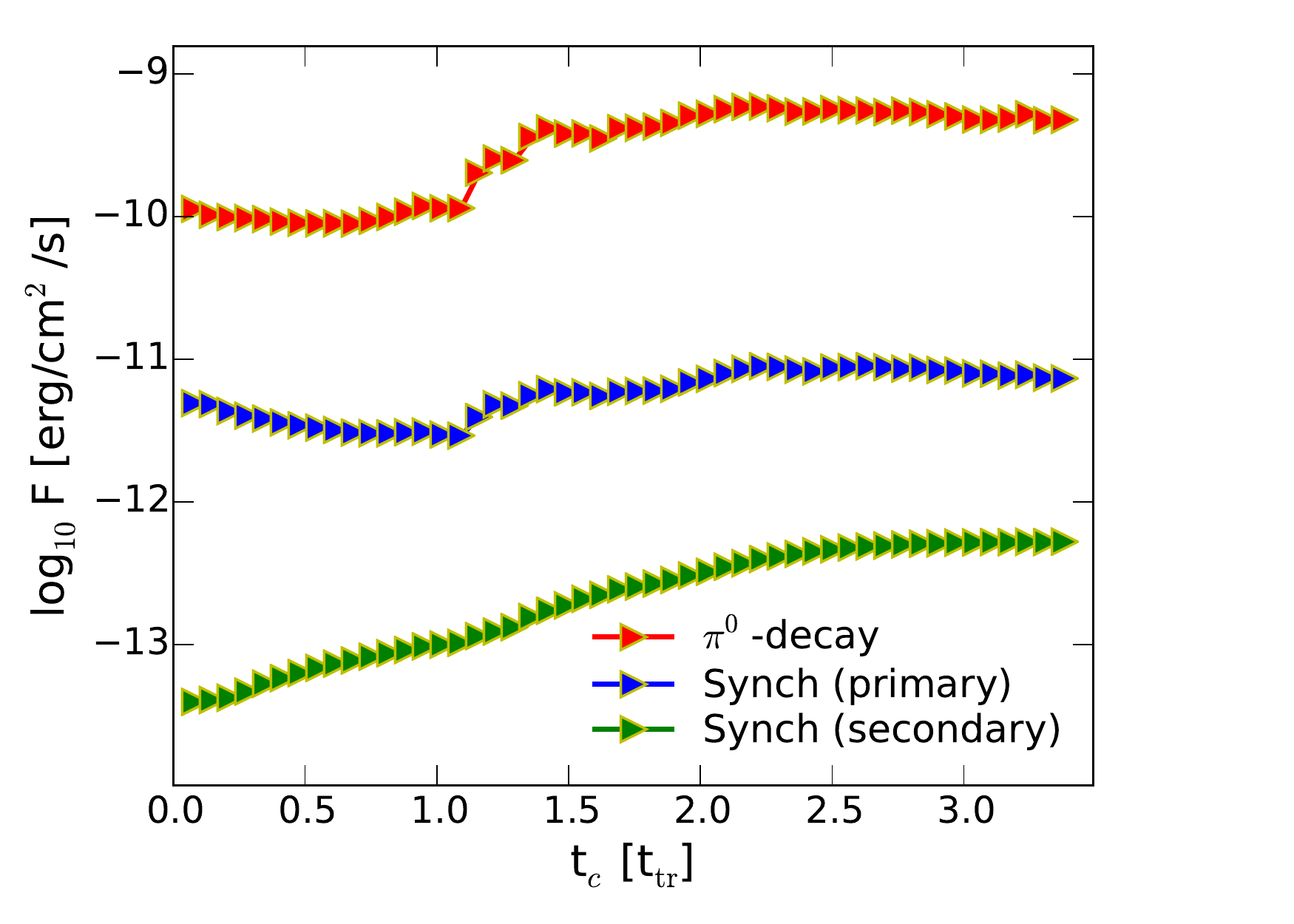}}
\caption{The same as Figure~\ref{SED_W44} but for the NLDSA scenario. 
An overall normalization factor of 0.05 is applied to the model spectra to explain the data.
}
\label{SED_W44_2}
\end{figure}

We perform the same simulation as the re-acceleration case and plot the resulting broadband SED in Figure~\ref{SED_W44_2}.   
In the results, we find a number of interesting differences with the former model: 

The first difference is that both the $\gamma$-ray and radio flux do not experience a boost after the transition of the shock to radiative phase as drastic as we witness in the previous case, only a factor of a few here versus a few 100 in the re-acceleration model. In this particular model, the instantaneous efficiency of DSA (i.e. fraction of incoming energy flux in the shock rest frame converted into relativistic particles), reaches 70 to 80$\%$ maximum during the early phase of the cloud shock evolution when the shock velocity is still high, and the pressure ratio between the non-thermal components (CR and magnetic) and the total (i.e. non-thermal plus thermal), i.e., $(P_\mrm{CR}+P_{B_\perp})/P_\mrm{tot}$, reaches about 0.7 downstream before the shock becomes radiative, and close to 1 after the gas has cooled down rapidly through radiating away its internal energy. 
As expected, the non-thermal emission are much stronger than the re-acceleration case in the initial phase. However,
we find that the high non-thermal pressure can support the radiatively cooling gas and prevent it from collapsing into a prominent dense shell. As a result, the effect from rapid compression is not as strong in this scenario. The radio-to-$\gamma$-ray flux ratio here is very similar to that in the re-acceleration model since the adopted $K_{ep} = 0.003$ is close to the number ratio between the Galactic CR electrons and protons specified in Equation~\ref{seed}. In this model, the non-thermal pressure is about equally shared by the CRs and $B_\perp$ with $P_\mrm{CR}/P_{B_\perp} \gtrsim 1$ compared to roughly $0.01$ in the re-acceleration model above. The post-shock plasma-$\beta$ before and after the radiative transition is $\sim 1$ and $\gsim 0.01$ respectively. The absolute values are much lower than the re-acceleration case due to the much larger CR-amplified $B$-field and lower shocked temperature as a result of efficient DSA in the early phase. The total fraction of the initial cloud shock energy attributed to CR acceleration is about 33\% at the end of simulation in this model, much larger than the `test-particle' re-acceleration model we discussed above.  

The second difference is that the shock becomes radiative at a much earlier time than the previous scenario (at $t_c \approx 0.5\ t_\mrm{tr}$ versus $1.2\ t_\mrm{tr}$). This can be easily explained by the significantly lower post-shock temperature at any given time due to a much larger portion of the shock kinetic energy being converted into non-thermal particles by efficient DSA. As a result, the shock does not have to decelerate to a velocity as low as when DSA is inefficient to achieve a downstream temperature at which the gas can cool down effectively through radiation. A similar result has also been obtained by \citet{Bykov2013}. This also means that when the shock became radiative, the shock was still strong enough to support NLDSA so that at that time $P_\mrm{CR}$ has already dominated the post-shock pressure to support the gas from collapsing drastically. The different timing of the transition to radiative phase is also reflected in the evolution of the forbidden lines; brightening of the lines in the NLDSA model occurs at a substantially earlier phase than the re-acceleration model (bottom panel in Figure~\ref{Emis_forb}). 

The third difference can be found in the radio spectrum. The calculated $\gamma$-ray spectrum provides a good fit with W44 data, but the spectral index of the radio synchrotron emission is substantially steeper ($\alpha \approx -0.53$ to $-0.54$) than the former model and typical observed values. This model has a much more interesting time evolution of the shock compression ratio as well as the CR-driven amplification of $B$-field in the precursor, which are both quickly decreasing with time as the shock decelerates in the dense medium; the compression ratio decreases from about 8 to below 4 from $t_c = 0$ to $3\ t_\mrm{tr}$, and even more dramatically for the amplified downstream $B$-field which drops from almost $1$~mG to only slightly higher than the compressed ambient field of the cloud. This results into the initial gradual decrease of both the $\pi^0$-decay $\gamma$-rays and synchrotron radio fluxes before the transition to the radiative phase. Although the well-known NLDSA effect of spectral hardening is witnessed for the accelerated electrons in the early stage, quickly DSA becomes inefficient and the newly accelerated electrons possess a spectral index steeper than $2$ in energy space. The formation of a prominent dense cool shell which compresses and re-energizes the downstream CRs like in the previous scenario fails to happen here as well. Consequently the overall electron spectrum is not hard enough to reproduce the observed radio index. This is even more true if we adopt a larger injection parameter (i.e., less efficient injection) and in such a case even the observed $\gamma$-ray spectrum cannot be explained due to a weaker $B$-field amplification and hence lower $p_\mrm{max}$ for the CR protons.      

The secondary-to-primary ratio of radio synchrotron emission is also higher throughout the evolution in this model. This can be explained by the fact that the previous model, which does not (re-)accelerate CRs so efficiently, relies on the formation of a dense cool shell later on in the evolution to boost the production rate of secondary leptons, while in this scenario protons are accelerated much more efficiently from the start to interact  with the more strongly compressed post-shock gas. For this reason as well as that discussed in the previous paragraph, the secondary flux also rises by a much larger factor in the re-acceleration scenario through the range of time shown.

An overall normalization factor of $0.05$ has to be applied to this model to reproduce the observed absolute fluxes, which is smaller than the former case and may not be easily attributable solely to the filling factor of the emission region. Since a prominent cool dense shell cannot be formed, the brightness profiles (along the direction perpendicular to the shock front) of the emission are also more diffuse and less filament-like than the former case. This is also true for the calculated optical and IR forbidden line emission.

\subsection{Beyond Spherical Symmetry and Other Aspects}

So far, we have investigated various interesting physical processes and their effects on the non-thermal emission at a cloud shock moving in a uniform dense medium, and compared them in the context of two particle acceleration scenarios for the emission mechanism. Since these discussions are based on models constructed by the one-dimensional \textit{CR-hydro-NEI} code, it is important to point out here several expected modifications of our results when the situation deviates from spherical symmetry. Although molecular clouds interacting with middle-aged SNRs are usually found to be more homogeneous than those interacting with young SNRs, discussion of these effects is nonetheless of significant interest. 

First of all, the molecular cloud can be clumpy and multi-phase. Although molecular cloud formation remains a highly active research field and many uncertainties still exist, one usually expects a lower density interclump medium surrounding some higher density gas clumps, most probably due to thermal instability and runaway cooling of the ISM \citep[e.g.,][]{Field1969, Koyama2000}. If the cloud is highly clumpy, it is possible that the cloud shock(s) cannot propagate in a dense medium for a duration sufficiently long to become fully radiative and to produce prominent dense cold shells. Also, shock-cloud interactions can happen at different times at different clouds such that, at a given SNR age, the integrated non-thermal emission becomes a superposition from multiple shocked clouds with distinct $t_c$, resulting in a change of the overall spectral properties. Other complexities like reflected shocks and MHD turbulences from shock-cloud interactions \citep[e.g.,][]{Inoue2012} can also happen. 

DSA at shocks in such a highly clumpy medium is expected to be modified compared to a spherically symmetric model, such as a modified CR escape process and magnetic field distribution. In the interclump gas with a lower density, cloud shocks can propagate at much faster speeds than in the dense clumps, so that possibly they can accelerate high-energy electrons capable of emitting localized non-thermal X-ray emission, as has been suggested to be the case for W44 \citep{Uchida2012}. In such a case, depending on the detailed cloud properties, $\gamma$-rays with a leptonic origin such as inverse-Compton scatterings may play a significant role in producing the observed high-energy emission. As the next extension of the current work, it is fruitful to perform a series of one-dimensional calculations to encompass both fast interclump shocks as well as the slower radiative shocks discussed here, which can be merged to realize a more complete representation of the situation in a clumpy cloud. In the further future, we foresee a generalization of our code to three-dimensions while maintaining it to be self-consistently coupled to DSA and CR transport.     

Finally, $\gamma$-ray emission from ``CR-illuminated'' clouds recently detected from the vicinity of middle-aged SNRs \citep[e.g.,][]{Uchiyama2012,Hanabata2014} has not been discussed here. A self-consistent model of such emission together with the shell component \citep[see e.g.,][for such models of young SNRs]{LKE2008,EB2011} is crucial for providing further constraints on the particle acceleration mechanism, such as the detailed processes of escape and propagation of CRs from the shock \citep[e.g.,][]{AharA1996,OhiraEscape2010}. Anisotropic diffusion is expected to play an important role in determining the $\gamma$-ray flux and morphology from these illuminated clouds \citep[e.g.,][]{MD2013,lara2013}. The inclusion of this component for middle-aged SNRs is another important next step for the work described in this paper. 

\section{Summary}
\label{summary}

We explored the possibility of explaining the bright non-thermal radio and GeV emission from middle-aged SNRs through a scenario with re-acceleration of Galactic CRs or NLDSA of thermally injected particles by a fast radiative cloud shock, under the assumption that a sufficiently strong magnetic turbulence is present in the cloud to support a Bohm-like diffusion. Using the \textit{CR-hydro-NEI} code, we followed the hydrodynamics of a shock propagating in a gas cloud with a density typical of those found in GeV-bright middle-aged SNRs, and at the same time calculated the time evolution of the (re-)accelerated CRs in the downstream and the generation of associated broadband non-thermal emission. 

In the re-acceleration scenario, we found that the transition of the cloud shock into the radiative phase, and the accompanying hydrodynamical effects on the post-shock conditions, play crucial roles in producing the bright non-thermal emission from the shell of these objects. The predicted general properties of the non-thermal emission agree well with observations and are illustrated by a comparison with the broadband spectrum of W44. The model produces strong oxygen lines such as [O I] and [O III], and the predicted surface brightness of [S II] agrees broadly with the observed value for W44, offering further support to the model. 

As for the NLDSA scenario with `thermal leakage' injection, a number of difficulties are found to explain observations. For example, we find that although the calculated $\gamma$-ray emission can reproduce the observed spectral characteristics quite well, the model fails to explain the radio continuum emission, with a spectral index too steep to reproduce typical observed values (e.g., W44 as well as others like IC 443 with similar indices), even though the model considered invokes a highly efficient injection for DSA, significant shock modification and hence spectral hardening. Models with less efficient injection are unable to explain observations due to even steeper radio indices and lower $\gamma$-ray cut-off energies. 

A number of other interesting differences between the behaviors of the two models are also discussed. For example, in the NLDSA model, the high non-thermal pressure in the post-shock gas is found to be an effective counter force to the rapid formation of a dense cool shell during the radiative cooling of the gas, therefore rapid compression plays a smaller role in generating the observed broadband emission compared to the re-acceleration case. The transition to the radiative phase also happens substantially earlier and is reflected by the time evolution of the  brightness of forbidden lines.

We have limited our discussion in this paper to the study of general properties of non-thermal emission from a one-dimensional but time-evolving cloud shock with self-consistent DSA. While our model is far from a complete account of the rich observed properties of any particular middle-aged SNR owing to its one-dimensional nature, we believe it succeeds to capture the essence of the evolutionary behavior of a radiative cloud shock, and the expected non-thermal emission produced which to first order agrees with radio continuum and $\gamma$-ray observations. This model can be considered an important first step towards a fuller understanding of non-thermal emission from these objects.   

\acknowledgements
The authors are grateful to the anonymous referee for offering helpful suggestions on improving the manuscript. 
SL and DCE express deep thanks to Andrei Bykov who provided valuable comments on the manuscript during a collaboration meeting at ISSI, Bern. SL acknowledges support from the JAXA International Top Young Fellowship, and the warm hospitality of the Harvard-Smithsonian Center for Astrophysics where a major part of this study was carried out under the SAO Visiting Scientist Program. DJP and POS acknowledge support from NASA contract NAS8-03060. DJP also acknowledges support from the Smithsonian Institution's Competitive Grants for Science Program. J.C.R. acknowledges support from grant HST-GO-13436. SN acknowledges support from the Japan Society for the Promotion of Science (Nos. 23340069, 24.02022, 25.03786 and 25610056). D.C.E acknowledges support from NASA grant NNX11AE03G. 

\bibliographystyle{aa} 
\bibliography{reference}

\begin{thebibliography}{79}
\expandafter\ifx\csname natexlab\endcsname\relax\def\natexlab#1{#1}\fi

\bibitem[{{Abdo} {et~al.}(2010{\natexlab{a}}){Abdo}, {Ackermann}, {Ajello},
  {Allafort}, {Baldini}, {Ballet}, {Barbiellini}, {Bastieri}, {Bechtol},
  {Bellazzini}, {Berenji}, {Blandford}, {Bloom}, {Bonamente}, {Borgland},
  {Bouvier}, {Brandt}, {Bregeon}, {Brigida}, {Bruel}, {Buehler}, {Buson},
  {Caliandro}, {Cameron}, {Caraveo}, {Carrigan}, {Casandjian}, {Cecchi}, {{\c
  C}elik}, {Chekhtman}, {Chiang}, {Ciprini}, {Claus}, {Cohen-Tanugi}, {Conrad},
  {Dermer}, {de Palma}, {Silva}, {Drell}, {Dubois}, {Dumora}, {Farnier},
  {Favuzzi}, {Fegan}, {Fukazawa}, {Fukui}, {Funk}, {Fusco}, {Gargano},
  {Gehrels}, {Germani}, {Giglietto}, {Giordano}, {Glanzman}, {Godfrey},
  {Grenier}, {Grove}, {Guiriec}, {Hadasch}, {Hanabata}, {Harding}, {Hays},
  {Horan}, {Hughes}, {J{\'o}hannesson}, {Johnson}, {Johnson}, {Kamae},
  {Katagiri}, {Kataoka}, {Kn{\"o}dlseder}, {Kuss}, {Lande}, {Latronico}, {Lee},
  {Lemoine-Goumard}, {Llena Garde}, {Longo}, {Loparco}, {Lovellette},
  {Lubrano}, {Makeev}, {Mazziotta}, {Michelson}, {Mitthumsiri}, {Mizuno},
  {Moiseev}, {Monte}, {Monzani}, {Morselli}, {Moskalenko}, {Murgia},
  {Nakamori}, {Nolan}, {Norris}, {Nuss}, {Ohno}, {Ohsugi}, {Omodei}, {Orlando},
  {Ormes}, {Ozaki}, {Panetta}, {Parent}, {Pelassa}, {Pepe}, {Pesce-Rollins},
  {Piron}, {Porter}, {Rain{\`o}}, {Rando}, {Razzano}, {Reimer}, {Reimer},
  {Reposeur}, {Rodriguez}, {Roth}, {Sadrozinski}, {Sander}, {Saz Parkinson},
  {Sgr{\`o}}, {Siskind}, {Smith}, {Smith}, {Spandre}, {Spinelli}, {Strickman},
  {Suson}, {Tajima}, {Takahashi}, {Takahashi}, {Tanaka}, {Thayer}, {Thayer},
  {Thompson}, {Tibaldo}, {Tibolla}, {Torres}, {Tosti}, {Uchiyama}, {Uehara},
  {Usher}, {Vasileiou}, {Vilchez}, {Vitale}, {Waite}, {Wang}, {Winer}, {Wood},
  {Yamamoto}, {Yamazaki}, {Yang}, {Ylinen}, \& {Ziegler}}]{AbdoEtalW282010}
{Abdo}, A.~A., {Ackermann}, M., {Ajello}, M., {et~al.} 2010{\natexlab{a}},
  \apj, 718, 348

\bibitem[{{Abdo} {et~al.}(2009){Abdo}, {Ackermann}, {Ajello}, {Baldini},
  {Ballet}, {Barbiellini}, {Baring}, {Bastieri}, {Baughman}, {Bechtol},
  {Bellazzini}, {Berenji}, {Blandford}, {Bloom}, {Bonamente}, {Borgland},
  {Bouvier}, {Bregeon}, {Brez}, {Brigida}, {Bruel}, {Burnett}, {Buson},
  {Caliandro}, {Cameron}, {Caraveo}, {Casandjian}, {Cecchi}, {{\c C}elik},
  {Chekhtman}, {Cheung}, {Chiang}, {Ciprini}, {Claus}, {Cohen-Tanugi},
  {Cominsky}, {Conrad}, {Cutini}, {Dermer}, {de Angelis}, {de Palma}, {Digel},
  {Dormody}, {Silva}, {Drell}, {Dubois}, {Dumora}, {Farnier}, {Favuzzi},
  {Fegan}, {Focke}, {Fortin}, {Frailis}, {Fukazawa}, {Funk}, {Fusco},
  {Gargano}, {Gasparrini}, {Gehrels}, {Germani}, {Giavitto}, {Giebels},
  {Giglietto}, {Giordano}, {Glanzman}, {Godfrey}, {Grenier}, {Grondin},
  {Grove}, {Guillemot}, {Guiriec}, {Hanabata}, {Harding}, {Hayashida}, {Hays},
  {Hughes}, {Jackson}, {J{\'o}hannesson}, {Johnson}, {Johnson}, {Johnson},
  {Kamae}, {Katagiri}, {Kataoka}, {Katsuta}, {Kawai}, {Kerr}, {Kn{\"o}dlseder},
  {Kocian}, {Kuss}, {Lande}, {Latronico}, {Lemoine-Goumard}, {Longo},
  {Loparco}, {Lott}, {Lovellette}, {Lubrano}, {Makeev}, {Mazziotta}, {McEnery},
  {Meurer}, {Michelson}, {Mitthumsiri}, {Mizuno}, {Moiseev}, {Monte},
  {Monzani}, {Morselli}, {Moskalenko}, {Murgia}, {Nakamori}, {Nolan}, {Norris},
  {Nuss}, {Ohsugi}, {Okumura}, {Omodei}, {Orlando}, {Ormes}, {Paneque},
  {Parent}, {Pelassa}, {Pepe}, {Pesce-Rollins}, {Piron}, {Porter}, {Rain{\`o}},
  {Rando}, {Razzano}, {Reimer}, {Reimer}, {Reposeur}, {Ritz}, {Rodriguez},
  {Romani}, {Roth}, {Ryde}, {Sadrozinski}, {Sanchez}, {Sander}, {Saz
  Parkinson}, {Scargle}, {Schalk}, {Sgr{\`o}}, {Siskind}, {Smith}, {Smith},
  {Spandre}, {Spinelli}, {Strickman}, {Suson}, {Tajima}, {Takahashi},
  {Takahashi}, {Tanaka}, {Thayer}, {Thayer}, {Thompson}, {Tibaldo}, {Tibolla},
  {Torres}, {Tosti}, {Tramacere}, {Uchiyama}, {Usher}, {Vasileiou}, {Venter},
  {Vilchez}, {Vitale}, {Waite}, {Wang}, {Winer}, {Wood}, {Yamazaki}, {Ylinen},
  \& {Ziegler}}]{AbdoEtalW51C2009}
{Abdo}, A.~A., {Ackermann}, M., {Ajello}, M., {et~al.} 2009, ApJL, 706, L1

\bibitem[{{Abdo} {et~al.}(2010{\natexlab{b}}){Abdo}, {Ackermann}, {Ajello},
  {Baldini}, {Ballet}, {Barbiellini}, {Baring}, {Bastieri}, {Baughman},
  {Bechtol}, {Bellazzini}, {Berenji}, {Blandford}, {Bloom}, {Bonamente},
  {Borgland}, {Bregeon}, {Brez}, {Brigida}, {Bruel}, {Burnett}, {Buson},
  {Caliandro}, {Cameron}, {Caraveo}, {Casandjian}, {Cecchi}, {{\c C}elik},
  {Chekhtman}, {Cheung}, {Chiang}, {Ciprini}, {Claus}, {Cognard},
  {Cohen-Tanugi}, {Cominsky}, {Conrad}, {Cutini}, {Dermer}, {de Angelis}, {de
  Palma}, {Digel}, {do Couto e Silva}, {Drell}, {Dubois}, {Dumora}, {Espinoza},
  {Farnier}, {Favuzzi}, {Fegan}, {Focke}, {Fortin}, {Frailis}, {Fukazawa},
  {Funk}, {Fusco}, {Gargano}, {Gasparrini}, {Gehrels}, {Germani}, {Giavitto},
  {Giebels}, {Giglietto}, {Giordano}, {Glanzman}, {Godfrey}, {Grenier},
  {Grondin}, {Grove}, {Guillemot}, {Guiriec}, {Hanabata}, {Harding},
  {Hayashida}, {Hays}, {Hughes}, {Jackson}, {J{\'o}hannesson}, {Johnson},
  {Johnson}, {Johnson}, {Kamae}, {Katagiri}, {Kataoka}, {Katsuta}, {Kawai},
  {Kerr}, {Kn{\"o}dlseder}, {Kocian}, {Kramer}, {Kuss}, {Lande}, {Latronico},
  {Lemoine-Goumard}, {Longo}, {Loparco}, {Lott}, {Lovellette}, {Lubrano},
  {Lyne}, {Madejski}, {Makeev}, {Mazziotta}, {McEnery}, {Meurer}, {Michelson},
  {Mitthumsiri}, {Mizuno}, {Monte}, {Monzani}, {Morselli}, {Moskalenko},
  {Murgia}, {Nakamori}, {Nolan}, {Norris}, {Noutsos}, {Nuss}, {Ohsugi},
  {Omodei}, {Orlando}, {Ormes}, {Paneque}, {Parent}, {Pelassa}, {Pepe},
  {Pesce-Rollins}, {Piron}, {Porter}, {Rain{\`o}}, {Rando}, {Razzano},
  {Reimer}, {Reimer}, {Reposeur}, {Rochester}, {Rodriguez}, {Romani}, {Roth},
  {Ryde}, {Sadrozinski}, {Sanchez}, {Sander}, {Parkinson}, {Scargle},
  {Sgr{\`o}}, {Siskind}, {Smith}, {Smith}, {Spandre}, {Spinelli}, {Stappers},
  {Stecker}, {Strickman}, {Suson}, {Tajima}, {Takahashi}, {Takahashi},
  {Tanaka}, {Thayer}, {Thayer}, {Theureau}, {Thompson}, {Tibaldo}, {Tibolla},
  {Torres}, {Tosti}, {Tramacere}, {Uchiyama}, {Usher}, {Vasileiou}, {Venter},
  {Vilchez}, {Vitale}, {Waite}, {Wang}, {Winer}, {Wood}, {Yamazaki}, {Ylinen},
  \& {Ziegler}}]{AbdoEtalW442010}
{Abdo}, A.~A., {Ackermann}, M., {Ajello}, M., {et~al.} 2010{\natexlab{b}},
  Science, 327, 1103

\bibitem[{{Abdo} {et~al.}(2010{\natexlab{c}}){Abdo}, {Ackermann}, {Ajello},
  {Baldini}, {Ballet}, {Barbiellini}, {Bastieri}, {Baughman}, {Bechtol},
  {Bellazzini}, {Berenji}, {Blandford}, {Bloom}, {Bonamente}, {Borgland},
  {Bregeon}, {Brez}, {Brigida}, {Bruel}, {Burnett}, {Buson}, {Caliandro},
  {Cameron}, {Caraveo}, {Casandjian}, {Cecchi}, {{\c C}elik}, {Chekhtman},
  {Cheung}, {Chiang}, {Cillis}, {Ciprini}, {Claus}, {Cohen-Tanugi}, {Cominsky},
  {Conrad}, {Cutini}, {Dermer}, {de Angelis}, {de Palma}, {Silva}, {Drell},
  {Drlica-Wagner}, {Dubois}, {Dumora}, {Farnier}, {Favuzzi}, {Fegan}, {Focke},
  {Fortin}, {Frailis}, {Fukazawa}, {Funk}, {Fusco}, {Gargano}, {Gasparrini},
  {Gehrels}, {Germani}, {Giavitto}, {Giebels}, {Giglietto}, {Giordano},
  {Glanzman}, {Godfrey}, {Grenier}, {Grondin}, {Grove}, {Guillemot}, {Guiriec},
  {Hanabata}, {Harding}, {Hayashida}, {Hughes}, {Jackson}, {J{\'o}hannesson},
  {Johnson}, {Johnson}, {Johnson}, {Kamae}, {Katagiri}, {Kataoka}, {Kawai},
  {Kerr}, {Kn{\"o}dlseder}, {Kocian}, {Kuss}, {Lande}, {Latronico}, {Lee},
  {Lemoine-Goumard}, {Longo}, {Loparco}, {Lott}, {Lovellette}, {Lubrano},
  {Madejski}, {Makeev}, {Mazziotta}, {Meurer}, {Michelson}, {Mitthumsiri},
  {Moiseev}, {Monte}, {Monzani}, {Morselli}, {Moskalenko}, {Murgia},
  {Nakamori}, {Nolan}, {Norris}, {Nuss}, {Ohsugi}, {Orlando}, {Ormes}, {Ozaki},
  {Paneque}, {Panetta}, {Parent}, {Pelassa}, {Pepe}, {Pesce-Rollins}, {Piron},
  {Porter}, {Rain{\`o}}, {Rando}, {Razzano}, {Reimer}, {Reimer}, {Reposeur},
  {Rochester}, {Rodriguez}, {Romani}, {Roth}, {Ryde}, {Sadrozinski}, {Sanchez},
  {Sander}, {Saz Parkinson}, {Scargle}, {Sgr{\`o}}, {Siskind}, {Smith},
  {Smith}, {Spandre}, {Spinelli}, {Strickman}, {Strong}, {Suson}, {Tajima},
  {Takahashi}, {Takahashi}, {Tanaka}, {Thayer}, {Thayer}, {Thompson},
  {Tibaldo}, {Torres}, {Tosti}, {Tramacere}, {Uchiyama}, {Usher}, {Van Etten},
  {Vasileiou}, {Venter}, {Vilchez}, {Vitale}, {Waite}, {Wang}, {Winer}, {Wood},
  {Ylinen}, \& {Ziegler}}]{AbdoEtalIC4432010}
{Abdo}, A.~A., {Ackermann}, M., {Ajello}, M., {et~al.} 2010{\natexlab{c}},
  \apj, 712, 459

\bibitem[{{Ackermann} {et~al.}(2013){Ackermann}, {Ajello}, {Allafort},
  {Baldini}, {Ballet}, {Barbiellini}, {Baring}, {Bastieri}, {Bechtol},
  {Bellazzini}, {Blandford}, {Bloom}, {Bonamente}, {Borgland}, {Bottacini},
  {Brandt}, {Bregeon}, {Brigida}, {Bruel}, {Buehler}, {Busetto}, {Buson},
  {Caliandro}, {Cameron}, {Caraveo}, {Casandjian}, {Cecchi}, {{\c C}elik},
  {Charles}, {Chaty}, {Chaves}, {Chekhtman}, {Cheung}, {Chiang}, {Chiaro},
  {Cillis}, {Ciprini}, {Claus}, {Cohen-Tanugi}, {Cominsky}, {Conrad}, {Corbel},
  {Cutini}, {D'Ammando}, {de Angelis}, {de Palma}, {Dermer}, {do Couto e
  Silva}, {Drell}, {Drlica-Wagner}, {Falletti}, {Favuzzi}, {Ferrara},
  {Franckowiak}, {Fukazawa}, {Funk}, {Fusco}, {Gargano}, {Germani},
  {Giglietto}, {Giommi}, {Giordano}, {Giroletti}, {Glanzman}, {Godfrey},
  {Grenier}, {Grondin}, {Grove}, {Guiriec}, {Hadasch}, {Hanabata}, {Harding},
  {Hayashida}, {Hayashi}, {Hays}, {Hewitt}, {Hill}, {Hughes}, {Jackson},
  {Jogler}, {J{\'o}hannesson}, {Johnson}, {Kamae}, {Kataoka}, {Katsuta},
  {Kn{\"o}dlseder}, {Kuss}, {Lande}, {Larsson}, {Latronico}, {Lemoine-Goumard},
  {Longo}, {Loparco}, {Lovellette}, {Lubrano}, {Madejski}, {Massaro}, {Mayer},
  {Mazziotta}, {McEnery}, {Mehault}, {Michelson}, {Mignani}, {Mitthumsiri},
  {Mizuno}, {Moiseev}, {Monzani}, {Morselli}, {Moskalenko}, {Murgia},
  {Nakamori}, {Nemmen}, {Nuss}, {Ohno}, {Ohsugi}, {Omodei}, {Orienti},
  {Orlando}, {Ormes}, {Paneque}, {Perkins}, {Pesce-Rollins}, {Piron}, {Pivato},
  {Rain{\`o}}, {Rando}, {Razzano}, {Razzaque}, {Reimer}, {Reimer}, {Ritz},
  {Romoli}, {S{\'a}nchez-Conde}, {Schulz}, {Sgr{\`o}}, {Simeon}, {Siskind},
  {Smith}, {Spandre}, {Spinelli}, {Stecker}, {Strong}, {Suson}, {Tajima},
  {Takahashi}, {Takahashi}, {Tanaka}, {Thayer}, {Thayer}, {Thompson},
  {Thorsett}, {Tibaldo}, {Tibolla}, {Tinivella}, {Troja}, {Uchiyama}, {Usher},
  {Vandenbroucke}, {Vasileiou}, {Vianello}, {Vitale}, {Waite}, {Werner},
  {Winer}, {Wood}, {Wood}, {Yamazaki}, {Yang}, \& {Zimmer}}]{Ackermann2013}
{Ackermann}, M., {Ajello}, M., {Allafort}, A., {et~al.} 2013, Science, 339, 807

\bibitem[{{Aharonian} \& {Atoyan}(1996)}]{AharA1996}
{Aharonian}, F.~A. \& {Atoyan}, A.~M. 1996, \aap, 309, 917

\bibitem[{{Bale} {et~al.}(2013){Bale}, {Pulupa}, {Salem}, {Chen}, \&
  {Quataert}}]{Bale2013}
{Bale}, S.~D., {Pulupa}, M., {Salem}, C., {Chen}, C.~H.~K., \& {Quataert}, E.
  2013, \apjl, 769, L22

\bibitem[{{Blandford} \& {Cowie}(1982)}]{BC82}
{Blandford}, R.~D. \& {Cowie}, L.~L. 1982, \apj, 260, 625

\bibitem[{{Blasi}(2004)}]{Blasi2004}
{Blasi}, P. 2004, APh, 21, 45

\bibitem[{{Blondin} {et~al.}(1998){Blondin}, {Wright}, {Borkowski}, \&
  {Reynolds}}]{Blondin1998}
{Blondin}, J.~M., {Wright}, E.~B., {Borkowski}, K.~J., \& {Reynolds}, S.~P.
  1998, \apj, 500, 342

\bibitem[{{Bykov} {et~al.}(2014){Bykov}, {Ellison}, {Osipov}, \&
  {Vladimirov}}]{Bykov2014}
{Bykov}, A.~M., {Ellison}, D.~C., {Osipov}, S.~M., \& {Vladimirov}, A.~E. 2014,
  \apj, 789, 137

\bibitem[{{Bykov} {et~al.}(2013){Bykov}, {Malkov}, {Raymond},
  {Krassilchtchikov}, \& {Vladimirov}}]{Bykov2013}
{Bykov}, A.~M., {Malkov}, M.~A., {Raymond}, J.~C., {Krassilchtchikov}, A.~M.,
  \& {Vladimirov}, A.~E. 2013, \ssr, 178, 599

\bibitem[{{Cardillo} {et~al.}(2014){Cardillo}, {Tavani}, {Giuliani},
  {Yoshiike}, {Sano}, {Fukuda}, {Fukui}, {Castelletti}, \&
  {Dubner}}]{Cardillo2014}
{Cardillo}, M., {Tavani}, M., {Giuliani}, A., {et~al.} 2014, \aap, 565, A74

\bibitem[{{Castelletti} {et~al.}(2007){Castelletti}, {Dubner}, {Brogan}, \&
  {Kassim}}]{Castelletti2007}
{Castelletti}, G., {Dubner}, G., {Brogan}, C., \& {Kassim}, N.~E. 2007, \aap,
  471, 537

\bibitem[{{Castro} \& {Slane}(2010)}]{cs10}
{Castro}, D. \& {Slane}, P. 2010, \apj, 717, 372

\bibitem[{{Chevalier}(1999)}]{Chevalier99}
{Chevalier}, R.~A. 1999, \apj, 511, 798

\bibitem[{{Chevalier} \& {Imamura}(1982)}]{CI1982}
{Chevalier}, R.~A. \& {Imamura}, J.~N. 1982, \apj, 261, 543

\bibitem[{{Cox}(1972)}]{Cox1972}
{Cox}, D.~P. 1972, \apj, 178, 143

\bibitem[{{Cox} {et~al.}(1999){Cox}, {Shelton}, {Maciejewski}, {Smith},
  {Plewa}, {Pawl}, \& {R{\'o}{\.z}yczka}}]{Cox1999}
{Cox}, D.~P., {Shelton}, R.~L., {Maciejewski}, W., {et~al.} 1999, \apj, 524,
  179

\bibitem[{{Cui} \& {Cox}(1992)}]{Cui1992}
{Cui}, W. \& {Cox}, D.~P. 1992, \apj, 401, 206

\bibitem[{{Draine} \& {McKee}(1993)}]{Draine1993}
{Draine}, B.~T. \& {McKee}, C.~F. 1993, \araa, 31, 373

\bibitem[{{Drury} {et~al.}(1996){Drury}, {Duffy}, \& {Kirk}}]{DDK96}
{Drury}, L.~O., {Duffy}, P., \& {Kirk}, J.~G. 1996, \aap, 309, 1002

\bibitem[{{Ellison} \& {Bykov}(2011)}]{EB2011}
{Ellison}, D.~C. \& {Bykov}, A.~M. 2011, \apj, 731, 87

\bibitem[{{Field} {et~al.}(1969){Field}, {Goldsmith}, \& {Habing}}]{Field1969}
{Field}, G.~B., {Goldsmith}, D.~W., \& {Habing}, H.~J. 1969, \apjl, 155, L149

\bibitem[{{Fujita} {et~al.}(2009){Fujita}, {Ohira}, {Tanaka}, \&
  {Takahara}}]{FujitaEtal2009}
{Fujita}, Y., {Ohira}, Y., {Tanaka}, S.~J., \& {Takahara}, F. 2009, ApJL, 707,
  L179

\bibitem[{{Ghavamian} {et~al.}(2007){Ghavamian}, {Laming}, \&
  {Rakowski}}]{GLR2007}
{Ghavamian}, P., {Laming}, J.~M., \& {Rakowski}, C.~E. 2007, ApJL, 654, L69

\bibitem[{{Ghavamian} {et~al.}(2001){Ghavamian}, {Raymond}, {Smith}, \&
  {Hartigan}}]{Ghavamian2001}
{Ghavamian}, P., {Raymond}, J., {Smith}, R.~C., \& {Hartigan}, P. 2001, \apj,
  547, 995

\bibitem[{{Giuliani} {et~al.}(2011){Giuliani}, {Cardillo}, {Tavani}, {Fukui},
  {Yoshiike}, {Torii}, {Dubner}, {Castelletti}, {Barbiellini}, {Bulgarelli},
  {Caraveo}, {Costa}, {Cattaneo}, {Chen}, {Contessi}, {Del Monte},
  {Donnarumma}, {Evangelista}, {Feroci}, {Gianotti}, {Lazzarotto}, {Lucarelli},
  {Longo}, {Marisaldi}, {Mereghetti}, {Pacciani}, {Pellizzoni}, {Piano},
  {Picozza}, {Pittori}, {Pucella}, {Rapisarda}, {Rappoldi}, {Sabatini},
  {Soffitta}, {Striani}, {Trifoglio}, {Trois}, {Vercellone}, {Verrecchia},
  {Vittorini}, {Colafrancesco}, {Giommi}, \& {Bignami}}]{Giuliani2011}
{Giuliani}, A., {Cardillo}, M., {Tavani}, M., {et~al.} 2011, \apjl, 742, L30

\bibitem[{{Gnat} \& {Sternberg}(2009)}]{Gnat2009}
{Gnat}, O. \& {Sternberg}, A. 2009, \apj, 693, 1514

\bibitem[{{Hanabata} {et~al.}(2014){Hanabata}, {Katagiri}, {Hewitt}, {Ballet},
  {Fukazawa}, {Fukui}, {Hayakawa}, {Lemoine-Goumard}, {Pedaletti}, {Strong},
  {Torres}, \& {Yamazaki}}]{Hanabata2014}
{Hanabata}, Y., {Katagiri}, H., {Hewitt}, J.~W., {et~al.} 2014, \apj, 786, 145

\bibitem[{{Helder} {et~al.}(2010){Helder}, {Kosenko}, \& {Vink}}]{Helder2010}
{Helder}, E.~A., {Kosenko}, D., \& {Vink}, J. 2010, \apjl, 719, L140

\bibitem[{{Hollenbach} \& {McKee}(1989)}]{HM89}
{Hollenbach}, D. \& {McKee}, C.~F. 1989, \apj, 342, 306

\bibitem[{{Inoue} {et~al.}(2012){Inoue}, {Yamazaki}, {Inutsuka}, \&
  {Fukui}}]{Inoue2012}
{Inoue}, T., {Yamazaki}, R., {Inutsuka}, S.-i., \& {Fukui}, Y. 2012, \apj, 744,
  71

\bibitem[{{Itoh} \& {Masai}(1989)}]{Itoh1989}
{Itoh}, H. \& {Masai}, K. 1989, \mnras, 236, 885

\bibitem[{{Kawasaki} {et~al.}(2002){Kawasaki}, {Ozaki}, {Nagase}, {Masai},
  {Ishida}, \& {Petre}}]{Kawasaki2002}
{Kawasaki}, M.~T., {Ozaki}, M., {Nagase}, F., {et~al.} 2002, \apj, 572, 897

\bibitem[{{Kimoto} \& {Chernoff}(1997)}]{Kimoto1997}
{Kimoto}, P.~A. \& {Chernoff}, D.~F. 1997, \apj, 485, 274

\bibitem[{{Koo} \& {Heiles}(1995)}]{Koo1995}
{Koo}, B.-C. \& {Heiles}, C. 1995, \apj, 442, 679

\bibitem[{{Koyama} \& {Inutsuka}(2000)}]{Koyama2000}
{Koyama}, H. \& {Inutsuka}, S.-I. 2000, \apj, 532, 980

\bibitem[{{Lee} {et~al.}(2010){Lee}, {Raymond}, {Park}, {Blair}, {Ghavamian},
  {Winkler}, \& {Korreck}}]{JJ2010}
{Lee}, J.-J., {Raymond}, J.~C., {Park}, S., {et~al.} 2010, \apjl, 715, L146

\bibitem[{{Lee} {et~al.}(2012){Lee}, {Ellison}, \& {Nagataki}}]{LEN2012}
{Lee}, S.-H., {Ellison}, D.~C., \& {Nagataki}, S. 2012, \apj, 750, 156

\bibitem[{{Lee} {et~al.}(2008){Lee}, {Kamae}, \& {Ellison}}]{LKE2008}
{Lee}, S.-H., {Kamae}, T., \& {Ellison}, D.~C. 2008, \apj, 686, 325

\bibitem[{{Lee} {et~al.}(2014){Lee}, {Patnaude}, {Ellison}, {Nagataki}, \&
  {Slane}}]{LPENS2014}
{Lee}, S.-H., {Patnaude}, D.~J., {Ellison}, D.~C., {Nagataki}, S., \& {Slane},
  P.~O. 2014, \apj, 791, 97

\bibitem[{{Malkov} {et~al.}(2011){Malkov}, {Diamond}, \& {Sagdeev}}]{MDS2011}
{Malkov}, M.~A., {Diamond}, P.~H., \& {Sagdeev}, R.~Z. 2011, Nature
  Communications, 2

\bibitem[{{Malkov} {et~al.}(2013){Malkov}, {Diamond}, {Sagdeev}, {Aharonian},
  \& {Moskalenko}}]{MD2013}
{Malkov}, M.~A., {Diamond}, P.~H., {Sagdeev}, R.~Z., {Aharonian}, F.~A., \&
  {Moskalenko}, I.~V. 2013, \apj, 768, 73

\bibitem[{{Mavromatakis} {et~al.}(2003){Mavromatakis}, {Boumis}, \&
  {Goudis}}]{MBG03}
{Mavromatakis}, F., {Boumis}, P., \& {Goudis}, C.~D. 2003, \aap, 405, 591

\bibitem[{{Medina} {et~al.}(2014){Medina}, {Raymond}, {Edgar}, {Caldwell},
  {Fesen}, \& {Milisavljevic}}]{Medina2014}
{Medina}, A.~A., {Raymond}, J.~C., {Edgar}, R.~J., {et~al.} 2014, \apj, 791, 30

\bibitem[{{Narayan} \& {Medvedev}(2001)}]{Narayan2001}
{Narayan}, R. \& {Medvedev}, M.~V. 2001, \apjl, 562, L129

\bibitem[{{Nava} \& {Gabici}(2013)}]{lara2013}
{Nava}, L. \& {Gabici}, S. 2013, \mnras, 429, 1643

\bibitem[{{Ohira} {et~al.}(2010){Ohira}, {Murase}, \&
  {Yamazaki}}]{OhiraEscape2010}
{Ohira}, Y., {Murase}, K., \& {Yamazaki}, R. 2010, \aap, 513, A17+

\bibitem[{{Ohira} {et~al.}(2011){Ohira}, {Murase}, \&
  {Yamazaki}}]{OhiraEtal2011}
{Ohira}, Y., {Murase}, K., \& {Yamazaki}, R. 2011, \mnras, 410, 1577

\bibitem[{{Ozawa} {et~al.}(2009){Ozawa}, {Koyama}, {Yamaguchi}, {Masai}, \&
  {Tamagawa}}]{Ozawa2009}
{Ozawa}, M., {Koyama}, K., {Yamaguchi}, H., {Masai}, K., \& {Tamagawa}, T.
  2009, \apjl, 706, L71

\bibitem[{{Patnaude} {et~al.}(2009){Patnaude}, {Ellison}, \& {Slane}}]{PES2009}
{Patnaude}, D.~J., {Ellison}, D.~C., \& {Slane}, P. 2009, \apj, 696, 1956

\bibitem[{{Patnaude} {et~al.}(2015){Patnaude}, {Lee}, {Slane}, {Badenes},
  {Heger}, {Ellison}, \& {Nagataki}}]{PL2015}
{Patnaude}, D.~J., {Lee}, S.-H., {Slane}, P., {et~al.} 2015, \apj, submitted

\bibitem[{{Patnaude} {et~al.}(2010){Patnaude}, {Slane}, {Raymond}, \&
  {Ellison}}]{PSRE2010}
{Patnaude}, D.~J., {Slane}, P., {Raymond}, J.~C., \& {Ellison}, D.~C. 2010,
  \apj, 725, 1476

\bibitem[{{Planck Collaboration} {et~al.}(2014){Planck Collaboration},
  {Arnaud}, {Ashdown}, {Atrio-Barandela}, {Aumont}, {Baccigalupi}, {Banday},
  {Barreiro}, {Battaner}, {Benabed}, {Benoit-L{\'e}vy}, {Bernard},
  {Bersanelli}, {Bielewicz}, {Bobin}, {Bond}, {Borrill}, {Bouchet}, {Brogan},
  {Burigana}, {Cardoso}, {Catalano}, {Chamballu}, {Chiang}, {Christensen},
  {Colombi}, {Colombo}, {Crill}, {Curto}, {Cuttaia}, {Davies}, {Davis}, {de
  Bernardis}, {de Rosa}, {de Zotti}, {Delabrouille}, {D{\'e}sert}, {Dickinson},
  {Diego}, {Donzelli}, {Dor{\'e}}, {Dupac}, {En{\ss}lin}, {Eriksen}, {Finelli},
  {Forni}, {Frailis}, {Franceschi}, {Galeotta}, {Ganga}, {Giard},
  {Giraud-H{\'e}raud}, {Gonz{\'a}lez-Nuevo}, {G{\'o}rski}, {Green}, {Gregorio},
  {Gruppuso}, {Hansen}, {Harrison}, {Hern{\'a}ndez-Monteagudo}, {Herranz},
  {Hildebrandt}, {Holmes}, {Huffenberger}, {Jaffe}, {Jaffe}, {Keih{\"a}nen},
  {Keskitalo}, {Kisner}, {Kneissl}, {Knoche}, {Kunz}, {Kurki-Suonio},
  {L{\"a}hteenm{\"a}ki}, {Lamarre}, {Lasenby}, {Lawrence}, {Leonardi},
  {Liguori}, {Lilje}, {Linden-V{\o}rnle}, {Lubin}, {Maino}, {Marshall},
  {Martin}, {Mart{\'{\i}}nez-Gonz{\'a}lez}, {Masi}, {Matarrese}, {Mazzotta},
  {Melchiorri}, {Mendes}, {Mennella}, {Migliaccio}, {Miville-Desch{\^e}nes},
  {Moneti}, {Montier}, {Morgante}, {Mortlock}, {Munshi}, {Murphy}, {Naselsky},
  {Nati}, {Noviello}, {Novikov}, {Novikov}, {Oppermann}, {Oxborrow}, {Pagano},
  {Pajot}, {Paladini}, {Pasian}, {Peel}, {Perdereau}, {Perrotta}, {Piacentini},
  {Piat}, {Pietrobon}, {Plaszczynski}, {Pointecouteau}, {Polenta}, {Popa},
  {Pratt}, {Puget}, {Rachen}, {Reach}, {Reich}, {Reinecke}, {Remazeilles},
  {Renault}, {Rho}, {Ricciardi}, {Riller}, {Ristorcelli}, {Rocha}, {Rosset},
  {Roudier}, {Rusholme}, {Sandri}, {Savini}, {Scott}, {Stolyarov}, {Sutton},
  {Suur-Uski}, {Sygnet}, {Tauber}, {Terenzi}, {Toffolatti}, {Tomasi},
  {Tristram}, {Tucci}, {Umana}, {Valenziano}, {Van Tent}, {Vielva}, {Villa},
  {Wade}, {Yvon}, {Zacchei}, \& {Zonca}}]{Planck2014}
{Planck Collaboration}, {Arnaud}, M., {Ashdown}, M., {et~al.} 2014, ArXiv
  e-prints: 1409.5746

\bibitem[{{Rakowski} {et~al.}(2008){Rakowski}, {Laming}, \&
  {Ghavamian}}]{Rakowski2008}
{Rakowski}, C.~E., {Laming}, J.~M., \& {Ghavamian}, P. 2008, \apj, 684, 348

\bibitem[{{Raymond} {et~al.}(1988){Raymond}, {Hester}, {Cox}, {Blair}, {Fesen},
  \& {Gull}}]{Raymond1988}
{Raymond}, J.~C., {Hester}, J.~J., {Cox}, D., {et~al.} 1988, \apj, 324, 869

\bibitem[{{Reach} \& {Rho}(1996)}]{Reach1996}
{Reach}, W.~T. \& {Rho}, J. 1996, \aap, 315, L277

\bibitem[{{Reach} \& {Rho}(2000)}]{Reach2000}
{Reach}, W.~T. \& {Rho}, J. 2000, \apj, 544, 843

\bibitem[{{Reach} {et~al.}(2005){Reach}, {Rho}, \& {Jarrett}}]{Reach2005}
{Reach}, W.~T., {Rho}, J., \& {Jarrett}, T.~H. 2005, \apj, 618, 297

\bibitem[{{Rho} {et~al.}(1994){Rho}, {Petre}, {Schlegel}, \&
  {Hester}}]{Rho1994}
{Rho}, J., {Petre}, R., {Schlegel}, E.~M., \& {Hester}, J.~J. 1994, \apj, 430,
  757

\bibitem[{{Scaife} {et~al.}(2007){Scaife}, {Green}, {Battye}, {Davies},
  {Davis}, {Dickinson}, {Franzen}, {G{\'e}nova-Santos}, {Grainge}, {Hafez},
  {Hobson}, {Lasenby}, {Pooley}, {Rajguru}, {Rebolo}, {Rubi{\~n}o-Martin},
  {Saunders}, {Scott}, {Titterington}, {Waldram}, \& {Watson}}]{Scaife2007}
{Scaife}, A., {Green}, D.~A., {Battye}, R.~A., {et~al.} 2007, \mnras, 377, L69

\bibitem[{{Shimizu} {et~al.}(2012){Shimizu}, {Masai}, \&
  {Koyama}}]{Shimizu2012}
{Shimizu}, T., {Masai}, K., \& {Koyama}, K. 2012, \pasj, 64, 24

\bibitem[{{Slane} {et~al.}(2014){Slane}, {Bykov}, {Ellison}, {Dubner}, \&
  {Castro}}]{SlaneEtal2014}
{Slane}, P., {Bykov}, A., {Ellison}, D.~C., {Dubner}, G., \& {Castro}, D. 2014,
  \ssr

\bibitem[{{Sun} {et~al.}(2011){Sun}, {Reich}, {Reich}, {Xiao}, {Gao}, \&
  {Han}}]{Sun2011}
{Sun}, X.~H., {Reich}, P., {Reich}, W., {et~al.} 2011, \aap, 536, A83

\bibitem[{{Tang} \& {Chevalier}(2014{\natexlab{a}})}]{Tang2014a}
{Tang}, X. \& {Chevalier}, R.~A. 2014{\natexlab{a}}, \apjl, 784, L35

\bibitem[{{Tang} \& {Chevalier}(2014{\natexlab{b}})}]{Tang2014b}
{Tang}, X. \& {Chevalier}, R.~A. 2014{\natexlab{b}}, ArXiv e-prints: 1410.7510

\bibitem[{{Tavani} {et~al.}(2010){Tavani}, {Giuliani}, {Chen}, {Argan},
  {Barbiellini}, {Bulgarelli}, {Caraveo}, {Cattaneo}, {Cocco}, {Contessi},
  {D'Ammando}, {Costa}, {De Paris}, {Del Monte}, {Di Cocco}, {Donnarumma},
  {Evangelista}, {Ferrari}, {Feroci}, {Fuschino}, {Galli}, {Gianotti},
  {Labanti}, {Lapshov}, {Lazzarotto}, {Lipari}, {Longo}, {Marisaldi},
  {Mastropietro}, {Mereghetti}, {Morelli}, {Moretti}, {Morselli}, {Pacciani},
  {Pellizzoni}, {Perotti}, {Piano}, {Picozza}, {Pilia}, {Pucella}, {Prest},
  {Rapisarda}, {Rappoldi}, {Scalise}, {Rubini}, {Sabatini}, {Striani},
  {Soffitta}, {Trifoglio}, {Trois}, {Vallazza}, {Vercellone}, {Vittorini},
  {Zambra}, {Zanello}, {Pittori}, {Verrecchia}, {Santolamazza}, {Giommi},
  {Colafrancesco}, {Antonelli}, \& {Salotti}}]{Tavani2010}
{Tavani}, M., {Giuliani}, A., {Chen}, A.~W., {et~al.} 2010, \apjl, 710, L151

\bibitem[{{Uchida} {et~al.}(2012){Uchida}, {Koyama}, {Yamaguchi}, {Sawada},
  {Ohnishi}, {Tsuru}, {Tanaka}, {Yoshiike}, \& {Fukui}}]{Uchida2012}
{Uchida}, H., {Koyama}, K., {Yamaguchi}, H., {et~al.} 2012, \pasj, 64, 141

\bibitem[{{Uchiyama} {et~al.}(2010){Uchiyama}, {Blandford}, {Funk}, {Tajima},
  \& {Tanaka}}]{uchiyamaea10}
{Uchiyama}, Y., {Blandford}, R.~D., {Funk}, S., {Tajima}, H., \& {Tanaka}, T.
  2010, ApJL, 723, L122

\bibitem[{{Uchiyama} {et~al.}(2012){Uchiyama}, {Funk}, {Katagiri}, {Katsuta},
  {Lemoine-Goumard}, {Tajima}, {Tanaka}, \& {Torres}}]{Uchiyama2012}
{Uchiyama}, Y., {Funk}, S., {Katagiri}, H., {et~al.} 2012, \apjl, 749, L35

\bibitem[{{van der Laan}(1962)}]{Laan1962}
{van der Laan}, H. 1962, \mnras, 124, 125

\bibitem[{{Verner} {et~al.}(1996){Verner}, {Ferland}, {Korista}, \&
  {Yakovlev}}]{Verner96}
{Verner}, D.~A., {Ferland}, G.~J., {Korista}, K.~T., \& {Yakovlev}, D.~G. 1996,
  \apj, 465, 487

\bibitem[{{White} \& {Long}(1991)}]{White1991}
{White}, R.~L. \& {Long}, K.~S. 1991, \apj, 373, 543

\bibitem[{{Yamaguchi} {et~al.}(2009){Yamaguchi}, {Ozawa}, {Koyama}, {Masai},
  {Hiraga}, {Ozaki}, \& {Yonetoku}}]{Yamaguchi2009}
{Yamaguchi}, H., {Ozawa}, M., {Koyama}, K., {et~al.} 2009, \apjl, 705, L6

\bibitem[{{Yan} {et~al.}(2012){Yan}, {Lazarian}, \& {Schlickeiser}}]{Yan2012}
{Yan}, H., {Lazarian}, A., \& {Schlickeiser}, R. 2012, \apj, 745, 140

\bibitem[{{Yoshiike} {et~al.}(2013){Yoshiike}, {Fukuda}, {Sano}, {Ohama},
  {Moribe}, {Torii}, {Hayakawa}, {Okuda}, {Yamamoto}, {Tajima}, {Mizuno},
  {Nishimura}, {Kimura}, {Maezawa}, {Onishi}, {Mizuno}, {Ogawa}, {Giuliani},
  {Koo}, \& {Fukui}}]{Yoshiike2013}
{Yoshiike}, S., {Fukuda}, T., {Sano}, H., {et~al.} 2013, \apj, 768, 179

\bibitem[{{Zhou} \& {Chen}(2011)}]{Zhou2011}
{Zhou}, P. \& {Chen}, Y. 2011, \apj, 743, 4

\bibitem[{{Zhou} {et~al.}(2014){Zhou}, {Safi-Harb}, {Chen}, {Zhang}, {Jiang},
  \& {Ferrand}}]{Zhou2014}
{Zhou}, P., {Safi-Harb}, S., {Chen}, Y., {et~al.} 2014, \apj, 791, 87

\end{thebibliography}

\end{document}